\documentstyle[11pt,aaspp4,epsf]{article}

\newbox\grsign \setbox\grsign=\hbox{$>$} \newdimen\grdimen \grdimen=\ht\grsign
\newbox\simlessbox \newbox\simgreatbox
\setbox\simgreatbox=\hbox{\raise.5ex\hbox{$>$}\llap
     {\lower.5ex\hbox{$\sim$}}}\ht1=\grdimen\dp1=0pt
\setbox\simlessbox=\hbox{\raise.5ex\hbox{$<$}\llap
     {\lower.5ex\hbox{$\sim$}}}\ht2=\grdimen\dp2=0pt
\def\simgreat{\mathrel{\copy\simgreatbox}}
\def\simless{\mathrel{\copy\simlessbox}}

\def\vol#1  {{{#1}{\rm,}\ }}
\def\aa{{A\&A}, }     %Astronomy & Astrophysics%
\def\aasup{{A\&AS}, } %A & A Supplements%

\def\aj{{AJ}, }  %Astronomical Journal%
\def\apj{{ApJ}, } %Astrophysical Journal%
\def\apjs{{ApJS}, } %Astrophysical Journal Supplements%
\def\sci{{Sci}, }   % Science
\def\pasp{{PASP}, }  %Publications of the Astronomical%
\def\mnras{{MNRAS}, } %Monthly Notices of the Royal%
\def\nat{{Nat}, }     %Nature%

\begin{document}

\title{ A Study of Nine High-Redshift Clusters of Galaxies :  \\
        III. HST\altaffilmark{1} Morphology of Clusters 0023+0423 \\
	and 1604+4304}

\author{Lori M. Lubin\altaffilmark{2,3}}
\affil{Observatories of the Carnegie Institution of Washington, 813 Santa
Barbara Street, Pasadena, CA 91101}
\affil{lml@astro.caltech.edu}

\author{Marc Postman}
\affil{Space Telescope Science Institute\altaffilmark{4}, 3700 San 
Martin Dr., Baltimore, MD 21218}
\affil{postman@stsci.edu}

\author{J. B. Oke}
\affil{Palomar Observatory, California Institute of Technology, Pasadena,
CA 91125}
\affil{and}
\affil{National Research Council Canada, Herzberg Institute of Astrophysics, 
Dominion Astrophysical Observatory, 5071 W. Saanich Road, Victoria,
BC V8X 4M6}
\affil{oke@dao.nrc.ca}

\author{Kavan U. Ratnatunga}
\affil{Department of Physics, Carnegie Mellon University, Wean Hall 5000,
Forbes Avenue,}
\affil{Pittsburgh, PA 15213}
\affil{kavan@astro.phys.cmu.edu}

\author{James E. Gunn}
\affil{Princeton University Observatory, Peyton Hall, Princeton, NJ 08544}
\affil{jeg@astro.princeton.edu}

\author{John G. Hoessel}
\affil{Washburn Observatory, University of Wisconsin--Madison, 475 N. Charter
Street,}
\affil{Madison, Wisconsin 53706}
\affil{hoessel@uwfpc.astro.wisc.edu}

\author{Donald P. Schneider}
\affil{Department of Astronomy and Astrophysics, Pennsylvania State University,
525 Davey Lab,}
\affil{University Park, PA 16802}
\affil{dps@astro.psu.edu}

\vskip 1 cm
\centerline{Submitted for publication in the {\it Astronomical Journal}}

\altaffiltext{1}{Based on observations with the NASA/ESA {\it Hubble
Space Telescope,} obtained at the Space Telescope Science Institute
(STScI), which is operated by the Association of Universities for
Research in Astronomy (AURA), Inc., under National Aeronautics and
Space Administration (NASA) Contract NAS 5-26555.}

\altaffiltext{2}{Hubble Fellow}

\altaffiltext{3}{Present Address : Palomar Observatory, California
Institute of Technology, Mail Stop 105-24, Pasadena, CA 91125}

\altaffiltext{4}{Space Telescope Science Institute is operated by the
Association of Universities for Research in Astronomy, Inc.,
under contract to the National Aeronautics and Space Administration.}

\vfill
\eject

\begin{abstract}

We present a detailed morphological analysis of the galaxy populations
in the first two clusters to be completed in an extensive
observational study of nine high-redshift clusters of galaxies (Oke,
Postman \& Lubin 1988). These two clusters, CL0023+0423 and
CL1604+4304, are at redshifts of $z = 0.84$ and $z = 0.90$,
respectively. The morphological studies are based on high-angular
resolution imagery taken with WFPC2 aboard the {\it Hubble Space
Telescope}. These data are combined with deep, ground-based $BVRI$
photometry and spectra taken with the Keck 10-meter telescopes. The
morphological classifications presented in this paper consist of two
parts. Firstly, we provide a quantitative description of the
structural properties of $\sim 600$ galaxies per cluster field using
the Medium Deep Survey automated data reduction and object
classification software (Griffiths et al.\ 1994; Ratnatunga, Ostrander
\& Griffiths 1997). This analysis includes the galaxy position,
photometry, and best-fit bulge+disk model. Secondly, for the brightest
subsample of $\sim 200$ galaxies per cluster field, we provide a more
detailed morphological description through a visual classification
based on the revised Hubble classification scheme (e.g.\ Sandage 1961;
Sandage \& Bedke 1994).

Based on these classifications, we have examined the general relation
between galaxy morphology and other photometric and spectral
properties. We find that, as expected, the elliptical and S0 galaxies
are redder, on average, than the spirals and irregulars. In addition,
there is a strong correlation between morphology and spectral type. Of
the galaxies that are visually classified as ellipticals, the majority
show K star absorption spectra which are typical of nearby, red
early-type galaxies; however, a few are actually blue compact galaxies
with spectra characterized by fairly strong, narrow emission
lines. Normal late-type galaxies typically have spectra with blue
colors and [\ion{O}{2}] emission, while the presence of strong
star-formation features, such as extremely high equivalent width
[\ion{O}{2}], ${\rm H\beta}$, and/or [\ion{O}{3}] emission, is always
accompanied by peculiar morphologies which suggest recent mergers or
interactions.

We have used the statistical distributions of cluster galaxy
morphologies to probe the overall morphological composition of these
two systems.  This analysis reveals that the two clusters contain very
different galaxy populations.  CL0023+0423 has a galaxy population
which is more similar to groups of galaxies and the field.  This
system is almost completely dominated by spiral galaxies. CL1604+4304,
however, has a morphological composition which is more typical of a
normal, present-day cluster; early-type galaxies comprise $\sim 76\%$
of all galaxies brighter than $M_{V} = -19.0 + 5~{\rm log}~h$ in the
central $\sim 0.5~h^{-1}~{\rm Mpc}$. The ratio of S0 galaxies to
ellipticals in this cluster is $1.7 \pm 0.9$, consistent with local
cluster populations. The morphological results support the conclusions
of the dynamical analysis presented in the second paper of this series
(Postman, Lubin \& Oke 1998). CL0023+0423 consists of two galaxy
groups which are separated by $\sim 2900~{\rm km~s^{-1}}$ in radial
velocity. CL1604+4304, on the other hand, has a velocity distribution
indicating that it is already well-formed and relaxed. The
morphological composition, velocity dispersion, and implied mass of
the CL1604+4304 system are consistent with an Abell richness class 2
or 3 cluster.

\end{abstract}


\newpage

\section{Introduction}

The study of the galaxy populations of rich clusters provides
important constraints on the formation mechanisms of both clusters and
galaxies. Present--day clusters show a distinct correlation between
the structure of the cluster and the galaxy population. Irregular,
open clusters, such as Virgo, are spiral--rich. These systems show no
single, central condensation, though the galaxy surface density is at
least five times as great as the surrounding field ($n_{\rm gal} >
30~h^{3}~{\rm galaxies~{Mpc}^{-3}}$). These clusters are often highly
asymmetric and have significant degrees of substructure.  Dense,
centrally concentrated clusters, such as Coma, contain predominantly
early--type galaxies in their cores (Abell 1958; Oemler 1974; Dressler
1980a,b; Postman \& Geller 1984).  These clusters have a single,
prominent concentration among the bright member galaxies and typically
display a high--degree of spherical symmetry, though this does not
preclude evidence of some substructure. Central densities can reach as
high as $10^{4}~h^{3}~{\rm galaxies~{Mpc}^{-3}}$ (e.g\ Bahcall 1975;
Dressler 1978). In these regions, spiral galaxies comprise less than
10\% of the cluster population, while elliptical (E) and S0 galaxies
make up 90\% or more of the population. The ratio of S0s to
ellipticals is typically S0/E $\sim 2$ (Dressler 1980a). The galaxy
content of clusters is part of the general morphology--density
relation of galaxies; as the local density increases, the fraction of
elliptical and S0 galaxies increases, while the fraction of spiral
galaxies decreases (Hubble 1936; Dressler 1980a,b; Postman \& Geller
1984).

Previous studies of clusters of galaxies at $z < 1$ have revealed
significant evolution in the morphology and the color of the cluster
members.  One of the most notable of these changes is the progressive
blueing of cluster's galaxy population with redshift, a trend first
observed by Butcher \& Oemler (1984). They found that the fraction of
blue galaxies in a cluster is an increasing function of redshift,
indicating that clusters at redshifts of $z \sim 0.5$ are
significantly bluer than their low--redshift counterparts. At
redshifts of $z \sim 0.4$, the fraction of blue galaxies is $\sim
20\%$.  Recent HST image data reveal that many of these blue galaxies
are either ``normal'' spirals or have peculiar morphologies, resulting
in non--elliptical fractions which are 3 to 5 times higher than the
average current epoch cluster (Dressler et al.\ 1994; Couch et al.\
1994; Oemler, Dressler \& Butcher 1997; Dressler et al.\ 1997).
 
Detailed photometric observations of other intermediate redshift ($z
\simless 0.4$) clusters have confirmed the original results of Butcher
\& Oemler (e.g.\ Millington \& Peach 1990; Luppino et al.\ 1991; Rakos
\& Schombert 1995).  Even though these clusters show an increased
fraction of blue galaxies, they still contain a population of E/S0s
distinguished by extremely red colors and a tight color--magnitude
(CM) relation (a ``red envelope''). Both the mean color and the CM
relation are consistent with that of present--day ellipticals (e.g.\
Sandage 1972; Butcher \& Oemler 1984; Arag\'on-Salamanca et al.\ 1991;
Luppino et al.\ 1991; Molinari et al.\ 1994; Dressler et al.\ 1994;
Smail, Ellis \& Fitchett 1994; Stanford, Eisenhardt \& Dickinson
1995).

At redshifts of $z \simgreat 0.4$, the red envelope has moved
bluewards with redshift (Arag\'on-Salamanca et al.\ 1993; Smail et
al.\ 1994; Rakos \& Schombert 1995; Oke, Gunn \& Hoessel 1996; Lubin
1996; Ellis et al.\ 1997; Stanford, Eisenhardt \& Dickinson 1997). At
$z \sim 0.9$, there are few cluster members with colors nearly as red
as present--day ellipticals. The color distribution of this
high-redshift elliptical population is relatively narrow, and the
trend is uniform from cluster to cluster; this suggests a homogeneous
population which formed within a narrow time span (e.g.\ Bower, Lucey
\& Ellis 1992a,b). Dickinson (1995) finds similar results in a cluster
of galaxies which is associated with the $z = 1.206$ radio galaxy 3C
324. The galaxies in this cluster exhibit a narrow, red locus in the
CM magnitude diagram. This branch is $\sim 0.6$ mag bluer than the
expected ``no--evolution'' value, though the intrinsic rms color
scatter is only 0.2 mag. The observed color trend for the red envelope
of galaxies in this data is consistent with passive evolution of an
old stellar population formed by a single burst of star formation at
redshifts of $z \simgreat 2$.  The reasonably small color scatter
would imply closely synchronized intra--cluster star formation (Bower
et al.\ 1992a,b; Arag\'on-Salamanca et al.\ 1993; Dickinson 1995;
Ellis et al.\ 1997; Stanford et al.\ 1997).

The high-resolution imaging of HST has been essential in understanding
the evolutionary processes occurring at intermediate redshifts (see
e.g.\ Abraham et al.\ 1996a). Morphological classifications can be
made on scales of $\sim 1~{\rm kpc}$, providing a direct comparison
with ground-based classifications of nearby galaxies. A comprehensive
survey of 10 clusters of galaxies at $z = 0.37 - 0.56$ has revealed a
significant change relative to local clusters in the composition and
behavior of the galaxy populations (Smail et al.\ 1997, hereafter S97;
Dressler et al.\ 1997, hereafter D97). The authors have visually
classified over 6000 galaxies based on the Revised Hubble Scheme used
to classify nearby galaxies (e.g.\ Sandage 1961; Sandage \& Bedke
1994). These classifications are used to quantify the morphological
composition of each cluster. Their results indicate that the
morphology--density relation is qualitatively similar to that in the
local universe in those intermediate redshift clusters which are
centrally-concentrated and compact; however, the relation is
non-existent in the loose, open clusters. Even so, all of the clusters
exhibit a roughly similar make-up of galaxy morphologies.  The
fraction of ellipticals is the same or larger than that in local
clusters; the S0 fraction, however, is $\sim 2-3$ times lower, with a
corresponding increase in the cluster spiral population. These
findings imply that the elliptical population is already in place by
$z \sim 0.5$, but a large fraction of the S0 galaxies are formed
between redshifts of $z \sim 0.5$ and $z = 0$ (D97). However, it
should be noted that these classifications are typically derived from
images which are not of comparable quality to the local data. Because
of such uncertainty, the observed evolution in the S0 population is
still in contention (e.g.\ Stanford et al.\ 1997; Andreon, Davoust \&
Helm 1997; Andreon 1998).

Because there appears to be significant evolution occurring between
redshifts of $z \sim 0.5$ and the present epoch, it is critical to
extend these detailed observations to even higher redshifts if we are
to understand the formation of galaxy morphology, as well as the
mechanisms and timescales of this evolution. Therefore, we have
undertaken an extensive observational program to study nine candidate
clusters of galaxies at redshifts of $z > 0.6$. The cluster sample was
chosen from the Gunn, Hoessel \& Oke (1986) survey and the Palomar
Distant Cluster Survey (PDCS; Postman et al.\ 1996). For each cluster,
we are in the process of obtaining deep $BVRI$ photometry from Keck
and deep $K$ photometry from the KPNO 4-meter, low-resolution spectra
from Keck, and high angular resolution imagery from HST. The
observations and data processing procedures of this survey are the
subject of the first paper in this series (Oke, Postman \& Lubin 1998;
hereafter Paper I).

The first two clusters to be completed in this observational program
are CL0023+0423 and CL1604+4304 at redshifts of $z = 0.84$ and $z =
0.90$, respectively (see Paper I). In this paper, we have used HST
images to undertake a detailed morphological analysis of the galaxy
populations in the central regions of these two clusters. The
reduction and analysis of the Keck $BVRI$ photometry and spectra of
the galaxies in these cluster fields are discussed in the second paper
of this series (Postman, Lubin \& Oke 1998; hereafter Paper
II). However, the galaxy parameters presented in Paper II are used in
this paper, specifically for a comparison with the morphological
properties.

In Sect.\ 2, we provide a brief description of the data. In Sects.\ 3
and 4, we describe the automated and visual galaxy classification
procedures used in this paper and present a comparison between the two
techniques.  In Sect.\ 5, we examine the morphologies of the galaxies
in the two cluster fields. This includes the relationship between
morphology and other galaxy properties, as well as the overall
distribution of morphologies in the cluster. A discussion and summary
of our conclusions are presented in Sects.\ 6 and 7. In the following
analyses, we have assumed $q_{0} = 0.1$ (e.g\ Carlberg et al.\ 1996)
and $H_{0} = 100~h~{\rm km~s^{-1}~Mpc^{-1}}$.

\section{The Data}

In this section, we discuss briefly those aspects of the data
acquisition and processing which are applicable to the following
analyses. For more details, the reader is referred to Papers I and II.

\subsection{The HST Observations}

All nine clusters in our sample have been or will be observed with
WFPC2 on HST (see Table 3 of Paper I). The HST imaging covers one
WFPC2 field-of-view ($\sim 150^{''} \times 150^{''}$) of the central
region of each cluster. At the redshifts of CL0023+0423 and
CL1604+4304, this corresponds to approximately $0.5~h^{-1}~{\rm Mpc}$.

The CL0023+0423 field was observed in the F702W filter ($R_{702}$) for
a total of 17.9 ksec in 1995.  These observations were conducted by
the authors.  The F702W filter was chosen because it was the most
efficient choice for imaging $z > 0.6$ clusters given the typical
spectral energy distribution of the cluster galaxies combined with the
relatively high quantum efficiency and broad bandpass of the filter.

The CL1604+4304 field was observed in the F814W filter ($I_{814}$) for
32.0 ksec in 1994 and for 32.0 ksec in 1995. These observations were
conducted by J.\ Westphal. The F814W filter was most likely chosen
because observations in this passband correspond roughly to $B$ in the
cluster rest-frame.  In the analysis of the CL1604+4304 field, we use
only the 1995 $I_{814}$ observations; the 1994 $I_{814}$ data are not
used as the pointing was off by one third of an arcmin. The 1995 data
consist of two individual 16.0 ksec pointings which are shifted
slightly ($\delta x = -2.02^{''}$, $\delta y = -0.86^{''}$) relative
to each other. As discussed below, we have examined the two 1995
pointings separately.

For a comparison with the Keck photometric data, we have adopted a
Vega-based (``Johnson'') magnitude system for the photometric
calibrations in the $R_{702}$ and $I_{814}$ filters (see also Holtzman
et al.\ 1995a). The zero points used in this paper were computed by
using the routine SYNPHOT in the STSDAS IRAF reduction package. They
are 22.38 and 21.67 ${\rm mag~DN^{-1}~sec^{-1}}$ for the $R_{702}$ and
$I_{814}$ bandpasses, respectively.  These zero points are based on
aperture photometry measurements made within a $6^{''}$ radius
aperture (or effectively infinity for a point source; see Holtzman et
al.\ 1995b).  Figures~\ref{00hst} and \ref{16hst} show the full WFPC2
images of CL0023+0423 and CL1604+4304, respectively.

The WFPC2 observations provide sufficient resolution to permit a
detailed description of the morphological characteristics of the
galaxies in the central regions of these clusters. To achieve this
goal, we have performed both a qualitative and quantitative study of
the morphological and structural properties of the galaxies. The HST
images of each cluster field have been processed through the Medium
Deep Survey (MDS; Griffiths et al.\ 1994) data reduction pipeline in
order to detect and quantify the structural parameters of extended
sources in the field.  This pipeline-processing procedure includes
calibration, warm pixel correction, and image stacking to remove
cosmic rays. For a complete description of the MDS reduction pipeline,
see Ratnatunga, Ostrander \& Griffiths (1997), hereafter referred to
as ROG.  The final, calibrated images from this reduction procedure
are used for both the automated and visual morphological
classifications presented in this paper.

The automated classifications are performed by MDS software
specifically designed to detect objects in WFPC2 images and to provide
best-fit bulge+disk models.  For the CL1604+4304 field, the automated
morphological analyses are initially performed on the data from the
first 1995 pointing (given the STScI dataset designation of
u2845). All of the galaxies detected, analyzed and classified in this
dataset are then re-examined in an identical manner using the data of
the second 1995 pointing (designated as dataset u2846). The two
independent analyses are compared in order to provide a valuable
consistency check of the automated classification procedures (see
Sect.\ 3.1.1). In addition, each individual pointing of CL1604+4304
observations can be directly compared to the CL0023+0423 observations
because they reach roughly equivalent depths.

We present all objects classified as non-stellar by the automated MDS
software that are above the completeness limit. The completeness limit
corresponds to $SNRIL > 1.5$, where $SNRIL$ is the decimal logarithm
of the integrated signal-to-noise ratio in a region around each object
which is 1-$\sigma$ above the estimated local sky. This limit
corresponds to a surface-brightness limit of $\mu_{702} \approx
25.8~\rm{mag~per~arcsec^{2}}$ and $\mu_{814} \approx
25.5~\rm{mag~per~arcsec^{2}}$ and a total magnitude limit of roughly
$R_{702} \simeq 26.3$ and $I_{814} \simeq 26.0$ for an object in the
CL0023+0423 and CL1604+4304 fields, respectively.  The resulting
catalogs contain 674 galaxies in the CL0023+0423 field and 559
galaxies in the CL1604+4304 field.

In addition, the brightest subsample of these galaxies have been
visually classified according to the Revised Hubble system of nearby
galaxies (e.g.\ Sandage 1961; Sandage \& Bedke 1994). For each cluster
observation, we estimate the total magnitude limit down to which we
can accurately classify galaxies by eye. The total magnitude of each
galaxy is defined as the total magnitude of the best-fit galaxy model
determined in the automated classification procedure (see Sect.\
3.1.1; ROG). For the CL0023+0423 field, this corresponds to a limiting
magnitude of $R_{702} = 24.7$ and a subsample of 209 galaxies. For the
CL1604+4304 field, this corresponds to a limiting magnitude of
$I_{814} = 24.3$ and a subsample of 205 galaxies. The median $SNRIL$
of the galaxies in these subsamples is $\sim 2.6$.  Finding charts for
the visually-classified galaxies are shown in Figures~\ref{00index}
and \ref{16index} for the two HST fields, respectively. The MDS
identification number is indicated at the position of each
galaxy. These figures can be used as overlays with the HST images
presented in Figures~\ref{00hst} and \ref{16hst}.

\subsection{The Keck Observations}

All of the ground-based optical observations, both broad band and
spectroscopic, were taken with the Low Resolution Imaging Spectrometer
(LRIS; Oke et al.\ 1995) on either the Keck I or the Keck II
telescopes. We briefly describe below the observations and data
reduction; however, the reader is referred to Papers I and II for a
complete account of these observations.

\subsubsection{The Photometry}

The photometric survey was conducted in four broad band filters,
$BVRI$, which match the Cousins system well. The response curves of
these filters are shown in Figure 1 of Paper I. In imaging mode, LRIS
covers a field-of-view of $6 \times 8$ arcmin, considerably larger
than the WPFC2 field-of-view (see Sect.\ 2.1). The Keck observations
have been calibrated to the standard Cousins-Bessell-Landolt (Cape)
system through exposures of a number of Landolt standard star fields
(Landolt 1992).  The FOCAS package (Valdes 1982) was used to detect,
classify, and obtain aperture and isophotal magnitudes for all objects
in the co-added $BVRI$ images.  For the color analysis presented in
this paper, we use aperture magnitudes computed in a circular aperture
with a radius of $3^{''}$. This corresponds to a physical radius of
$\{14.60~14.92\}~h^{-1}~{\rm kpc}$ at $z = \{0.84~ 0.90\}$, the
redshifts of CL0023+0423 and CL1604+4304, respectively.  The limiting
magnitudes are $B = 25.1$, $V = 24.1$, $R = 23.5$, and $I = 21.7$ for
a 5-$\sigma$ detection in our standard aperture (for more details, see
Sects.\ 3.1 and 4.1 of Paper I).  Tables of the Keck photometry for
these clusters are given in Paper II.

Each galaxy that has been visually classified in our subsample of
brightest HST galaxies (see Sect.\ 2.1) is matched with the
corresponding galaxy from the ground-based observations. The typical
FWHM seeing in the LRIS imaging data is 0.96 arcsec, compared to the
0.1 arcsec resolution of HST.  In some cases, an individual galaxy in
the LRIS image is associated with more than one galaxy in the
corresponding HST image (see Sect.\ 5.1). We have not tried to obtain
ground-based photometry for the full sample of galaxies analyzed
through the MDS automated object classification as, at these faint
levels, the broader PSF of the LRIS images makes the
cross-identification with objects in the WFPC2 images unreliable.

\subsubsection{The Redshifts}

Multi-slit observations of the cluster field were made with LRIS in
spectroscopic mode using an $300~{\rm g~mm^{-1}}$ grating blazed at
5000 \AA. The chosen grating provided a dispersion of 2.35 \AA\ per
pixel and a spectral coverage of 5100 \AA. In order to obtain the full
wavelength range along the dispersion axis, the field-of-view of the
spectral observations was reduced from that of the imaging mode to
approximately $2 \times 8$ arcmin. For each cluster field, six
different slitmasks were made with approximately 35 objects per
mask. The exposure time for each mask was 1 hour. In practice, about
170 spectra were taken per cluster field, with $\sim 120$ yielding
measurable redshifts. Because the contamination rate is so large at
our cluster redshifts, most of these galaxies turn out to be field
galaxies, rather than cluster members (see Paper II).  For more
details on the observations and data reduction procedures, see Sects.\
3.2 and 4.2 of Paper I. Sample spectra from these fields are presented
in Figures~\ref{00spcluster} and \ref{16spcluster} (see also in
Figures 5, 20 and 21 of Paper II).

Because a large number of slitmasks is required to completely cover a
small area on the sky, the spectroscopic sample within the HST
field-of-view is very limited. Redshifts were measured for only 41 and
29 galaxies in the WFPC2 field-of-view of CL0023+0423 and CL1604+4304,
respectively (see also Sect.\ 5). Of the galaxies for which we have
taken spectra in these central regions, we have successfully measured
redshifts for $\sim 60\%$ of them in CL0023+0423 and $\sim 73\%$ of
them in CL1604+4304.  There are 12 confirmed cluster members in each
HST field.  We have determined cluster membership based on the
analysis of the velocity histograms in Sect.\ 3 of Paper II.

\section{Morphological Classification of Galaxies in the HST Data}

Below we present a detailed description of the automated and visual
techniques used to morphologically classify all of the galaxies (both
cluster and field) detected within the HST images of the two clusters.

\subsection{Automated Classifications}

\subsubsection{The MDS Analysis}

For a quantitative analysis of the galaxy properties in our two
cluster fields, we have processed our WFPC2 images through the data
processing procedures and the automated object detection and
classification algorithms designed by the Medium Deep Survey team. In
this paper, we present only a brief description of the automated
object classification procedures used to quantify the galaxy
morphology. More detailed information on the entire reduction and
identification process can be found in ROG and Ratnatunga et al.\
(1994, 1995), and at the MDS website address
http://astro.phys.cmu.edu/mle/index.html.

The final step in the MDS processing pipeline is the automated object
classification. This process involves a two-dimensional maximum
likelihood estimator (MLE) analysis which automatically optimizes the
model and the number of parameters to be fit to each object image. For
most of the galaxies, a 64--pixel square region around the center of
galaxy is examined. For larger galaxies, a 128--pixel square region is
chosen. In this selected region, a contour around each object which is
1-$\sigma$ above the estimated local sky is determined. The total
integrated signal-to-noise ratio of this region is a good measure of
the information content of the object image.  The completeness limit
of the object finding algorithm is $SNRIL \sim 1.5$; however, only
those galaxies with $SNRIL > 2$ have enough signal to be reliably fit
to the full two-component model discussed below (see ROG).  Therefore,
though we present all of the galaxies down to the completeness limit,
only those galaxies with $SNRIL > 2$ are used in the analysis
presented in this section.

Two scale-free, axisymmetric models are chosen to describe the galaxy
profiles. Elliptical galaxies are assumed to have a bulge-like, de
Vaucouleurs profile, while disk galaxies have a disk-like, exponential
profile.  Each profile is characterized by a major axis half--light
radius and axis ratio. Stellar (point-like) objects are examined
through the same procedure as the galaxy images, except that a
Gaussian profile is adopted. A maximum likelihood parameter estimation
procedure is used to determine the best model and the parameter
values. For each set of model parameters, a model image of the object
is created and compared with the actual object image. The likelihood
function is defined as the product of the probabilities for each model
pixel value with respect to the observed pixel value and its error
distribution. Finally, a best-fit model and its parameters are
determined for each object with the following classifications: bulge
$+$ disk, disk, bulge, galaxy (if the classification as disk or bulge
is not significant), stellar, or object (if there is no preference
between star and galaxy). For the details on the parameter fittings
and the maximum likelihood estimator, see ROG and the references
listed above.

All objects that are classified as non-stellar with a reliable
signal-to-noise ratio ($SNRIL > 1.5$) are listed in Table 1. Tables
1a--b list the relevant parameters of these fittings for the galaxies
in the CL0023+0423 and CL1604+4304 fields, respectively. These
parameters include the best-fit model, centroid, total magnitude,
half-light radius, orientation, axis ratios, bulge/(disk+bulge)
luminosity ratio, and bulge/disk half-light radius ratio. Table 1c
give notes on the parameters in Table 1a--b. Because of the size of
these tables, we have chosen not to publish Tables 1a--b; instead, a
machine-readable form of these tables can be obtained from the authors
L.\ Lubin or M.\ Postman and from the website
http://landru.stsci.edu:5000/hizclus/ftp.html (see also Note to Table
1c).

For CL1604+4304, the galaxy parameters presented in Table 1b and used
in the subsequent sections are obtained from the MDS analysis of the
first dataset (u2845) of these observations. The second dataset
(u2846) is used only as a test of the robustness of the MDS automated
model fitting (see Sect.\ 2.1).  Using the u2846 dataset, we have
re-examined in an identical manner all the galaxies analyzed in the
u2845 dataset and made a comparison of the model parameters of the
same galaxy derived from the two datasets. For this comparison, we use
those galaxies with $SNRIL > 2$ as only they have enough signal to be
reliably fit to the full two component model (see ROG). This
corresponds to a sample of 293 galaxies (including both cluster and
field).

Figure~\ref{mdscomp} presents these results for the best-fit model,
the total magnitude $m_{tot}$, the logarithm of the half-light radius
$R_{\frac{1}{2}}$, the disk ${(\frac{b}{a})}_{D}$ and bulge
${(\frac{b}{a})}_{B}$ axial ratios, and the bulge/(disk+bulge)
luminosity ratio $\frac{B}{D+B}$. The best-fit model, on average, is
the same in both datasets; however, the scatter between components is
large with 23\% of the objects having a different classification in
the two datasets. The median and standard deviation (as measured from
the interquartile range) of the ratio of $m_{tot}$ in the u2845 and
u2846 datasets is $1.00 \pm 0.01$, while the ratio of
$R_{\frac{1}{2}}$ in the two datasets is $1.02 \pm 0.19$. The medians
of both the disk and bulge axial ratios are 1. In addition, the median
of the bulge/(disk+bulge) luminosity ratio is 1, with 66\% of all
galaxies having an identical value in the second dataset. However, the
scatter and the resulting pattern in these three parameters (shown in
the bottom panels of Figure~\ref{mdscomp}) reflect the fact that, in
only 77\% of the cases, the individual galaxy has the same
classification in both datasets.  For the brighter subsample used in
the visual classifications (see Sect.\ 3.2), the $SNRIL$ is higher
with a median value of $\sim 2.6$ and, therefore, the comparison is
slightly better (see also ROG). The medians for all of the parameters
mentioned above are consistent with 1, typically with smaller
scatter. In 79\% of these cases, the galaxy has the same
classification (either disk, bulge, or disk+bulge) in both datasets.

\subsubsection{The Asymmetry Parameter}

In addition to the automated disk+bulge MDS analysis, we have also
measured a simple quantitative index of visual asymmetry ($A$) in each
galaxy. The asymmetry parameter has been used previously in
morphological studies of the galaxies in the Medium Deep Survey
(Abraham et al.\ 1996a) and the Hubble Deep Field (Abraham et al.\
1996b; van den Bergh et al.\ 1996). A similarly defined parameter is
also used by S97 in the morphological analysis of 10 intermediate
redshift clusters of galaxies. Here, we define the parameter in the
identical manner to Abraham et al.\ (1996a).

We use for this procedure the galaxy region which was examined in the
MDS analysis (see Sect.\ 3.1.1) which corresponds to an isophotal area
enclosed by pixels 1-$\sigma$ above the sky level. The asymmetry
parameter $A$ is determined by rotating the individual galaxy image
180 degrees about its center and subtracting the rotated image from
the original. The center of rotation is defined as the pixel with the
maximum value in the original galaxy image after it has been smoothed
with a Gaussian kernel of $\sigma = 1$ pixel. The parameter $A$ is
then defined as

\begin{equation}
A = {\frac{1}{2}}{\frac{\left| I_{R} - I_{0}\right|}{I_{0}}}
\end{equation}

\noindent where $I_{R}$ and $I_{0}$ is the total light (after sky
subtraction) in the rotated image and the original image,
respectively. Because the absolute value of the residual light is used
in this definition, noise in the images manifests itself as a positive
signal in $A$ even in perfectly symmetrical objects. Therefore, it is
necessary to remove this contribution by subtracting a small
correction factor from each measured asymmetry parameter. This
correction factor is simply the value of $A$ for a portion of the sky
with an area equal to that of the galaxy region. Typical values of the
asymmetry parameter range from $0.0$ to $\sim 0.6$ (see
Figure~\ref{arescomp}). Tables 1a--b include the asymmetry parameter
$A$ for each galaxy analyzed in the MDS automated analysis (see Sect.\
3.1.1).

\subsection{Visual Classifications}

Visual galaxy classification in the nearby universe has had a
considerably impact on theories of galaxy formation, environment, and
evolution. In light of this, we have classified by eye the brightest
subsample of galaxies in our two cluster fields. Unfortunately, this
can be an inherently uncertain and subjective process even locally
(for a nice demonstration, see Naim et al.\ 1995a,b). At high
redshift, the situation becomes even more complicated. Here, we must
contend with limited resolution, low signal-to-noise, $(1+z)^{4}$
cosmological dimming, and a $k$-correction which implies that we are
moving into the relatively poorly observed $U$ band (see e.g.\ Bohlin
et al.\ 1991; O'Connell 1997; Hibbard \& Vacca 1997). Detailed studies
of how morphological classifications of high-redshift galaxies differ
from ground-based observations of nearby systems and the resulting
observational biases are currently being made (e.g.\ Bershady et al.\
1994; Odewahn et al.\ 1996). For this visual study, we present the
classifications as is, with the strong caveat that these morphologies
can be used to draw conclusions based only on the most broad
differences in galaxy type.

For each individual HST observation, we estimate the total magnitude
limit to which we can still accurately classify galaxies by eye. For
the CL0023+0423 field, this corresponds to $R_{702} = 24.7$ or a
subsample of 209 galaxies. For the CL1604+4304 field, this corresponds
to $I_{814} = 24.3$ or a subsample of 205 galaxies (see Sect.\
2.1). Galaxies which have been classified as ``stellar'' in the MDS
automated analysis have been excluded. For the specifics of the galaxy
classifications, we have adopted the classification procedure used by
S97 to morphologically classify galaxies in intermediate redshift
clusters. This galaxy classification includes four components : (1)
the Revised Hubble type, (2) disturbance index -- the perceived
asymmetry of the galaxy image, (3) dynamical state -- the
interpretation of the cause of any observed disturbance, and (4)
comments.

For the first component, we have based our visual classifications on
the Revised Hubble scheme (e.g.\ Sandage 1961; Sandage \& Bedke 1994).
Here, early-type galaxies or spheroids are classified as either
ellipticals (E) or S0s. No finer subdivisions are employed because of
the limited resolution and modest signal-to-noise in the majority of
cases. Because of the difficulty of distinguishing between faint
ellipticals and S0 galaxies on CCD images, we have used the classes
E/S0 and S0/E to specify those cases which are ambiguous. The order
reflects which galaxy type is thought to be more likely. Spirals are,
in principle, assigned half-classes (e.g.\ Sab, Sbc, Scd); however, as
noted above, there is normally insufficient information for these
delineations to be extremely accurate. Galaxies with obvious bars are
indicated with a ``B'' (e.g.\ SBa) though, again, this specification
is not meant to be inclusive or an accurate measure of the true
fraction of barred galaxies in these fields.

In agreement with many previous authors (e.g.\ Couch et al.\ 1994;
Dressler et al.\ 1994; Griffiths et al.\ 1994; Cowie, Hu \& Songaila
1995; Abraham et al.\ 1996a; Lavery et al.\ 1996; Naim et al.\ 1997;
Brinchmann et al.\ 1997), we also find a large number of galaxies
which are unlike any nearby galaxy classified according to the Revised
Hubble Scheme. We have adopted two new classifications which can, for
the most part, encompass these galaxies. Firstly, as in S97, those
galaxies whose images are too small for a reliable classification are
indicated by an ``X'' for compact. These galaxies can be almost
stellar (or point-like) in appearance, but they have been classified
as galaxies in the MDS automated analysis (see Sect.\ 3.1.1). In some
cases, the ``X'' classification may be similar to the ``N''
classification which describes a class of galaxies which are
intermediate between Seyfert galaxies and quasars. Such galaxies have
been observed to have a centrally peaked blue source imposed on a
faint, extended red component (see e.g.\ Sandage 1973; Morgan \&
Dreiser 1983). However, this classification would also describe very
distant galaxies where cosmological dimming means that only the
high-surface-brightness bulges are visible. Secondly, those galaxies
with extremely peculiar shapes are indicated by a ``P'' for
peculiar. This class includes galaxies which appear severely deformed
due to interactions or tidal forces; galaxies with double nuclei or
other evidence of a merger; ``amorphous'' galaxies which appear as a
high-surface-brightness, smooth, unresolved sheet of light (Sandage \&
Brucato 1979); and galaxies which simply cannot be classified in the
traditional manner.  (Examples of galaxies in these classes can be
found in Figures~\ref{00hstz} and \ref{16hstz}.)

The second component is the disturbance index $D$ (S97). This
parameter ranges from 0 to 4 with the following definitions : 0 --
little or no asymmetry, 1 or 2 -- moderate or strong asymmetry, and 3
or 4 -- moderate or strong distortion.  This classification was
intended to be objective in that it is independent of the possible
reason of the disturbance.

The third component is the ``dynamical state.'' This parameter was
intended to be a subjective and interpretive judgment of the cause of
the disturbance and should, therefore, be viewed only as an educated
guess. The classes assigned to this parameter are : I -- tidal
interaction with a neighbor, M -- tidal interaction suggesting a
merger, T -- tidal feature without obvious cause, and C --
chaotic. For example, galaxies which were clearly disturbed or which
had two nuclei in a common envelope were designated as M for
``mergers.''  (See S97 for an additional discussion of these
classification procedures.) Finally, in the comments, we list the
relevant details on the morphology and the nearby surroundings of each
galaxy.

All of the galaxies in our HST images were classified independently on
a video display by L.\ Lubin and M.\ Postman (see Sect.\ 4.4.3 of
Paper I); in addition, expert identifications of all galaxies were
provided by Allan Sandage. Most importantly, he reviewed the rather
tricky separation between E and S0 galaxies. The independent
identifications of the three classifiers were merged by L.\ Lubin. In
most cases, the classifications agree to one class or better. That is,
there was agreement within one Hubble class for more than 75\% of all
galaxies which were visually classed. For the brightest 25\% of each
sample, this agreement improves to $\sim 90\%$.  The comments listed
in Table 2 describe those features which have affected the final
classification, as well as those features which may have been the
subject of a disagreement between the classifiers. In short, the
visual morphological classifications provide a good, general
indication of the class of the galaxy.  Tables 2a--b list the full
morphological information described above for the CL0023+0423 and
CL1604+4304 cluster fields, respectively. Table 2c gives notes to the
parameters in Tables 2a--b.

\section{Comparison between Visual and Automated Typing}

We use the brightest subsample of galaxies in the CL0023+0423 and
CL1604+4304 fields to compare the parameters of the visual
classifications (Sect.\ 3.2) with those from the automated,
algorithm-based classifications (Sect.\ 3.1). Because of the inherent
difficulties in performing visual classifications at these redshifts,
we use only the general classes of elliptical (E), S0, spiral (Sp),
and irregular/peculiar (Irr/Pec) in the following comparisons, even
though we have visually typed on a finer scale. Note that this sample
of galaxies includes both field and cluster galaxies.

First, we compare the galaxy's Hubble type with the best-fit model
from the MDS automated classification procedure for the combined
fields of CL0023+0423 and CL1604+4304. Table 3 shows the percentages
of galaxies visually classified as E, S0, Sp, or Irr/Pec which have
automated classifications of disk (D), bulge (B), or disk+bulge
(D+B). The sample contains a total of 44 E, 54 S0, 140 Sp, and 136
Irr/Pec galaxies. In 62\% of the cases, galaxies classified by eye as
spheroids (E or S0) have an indication of a bulge component (either B
or D+B) in the MLE fittings; however, in 38\% of the cases, a pure
disk is the best-fit model. For elliptical and S0 galaxies, 32\% and
43\%, respectively, are classified as a pure disk (see Table 3). The
large fraction of pure disk classifications may, in part, be caused by
the inability of an $r^{\frac{1}{4}}$ law to describe all
ellipticals. Studies of early-type galaxies in nearby clusters, from
brightest cluster galaxies (BCGs) to dwarf ellipticals, indicate that
in many cases a modified Hubble law (Oemler 1976) or a pure
exponential profile is more appropriate (e.g.\ Binggeli, Sandage \&
Tarenghi 1984; Schombert 1986, 1987; Burkert 1993; Coan et al.\ 1993;
Graham et al.\ 1996).

In particular, there appears to be a correlation between the power-law
$n$ in the generalized de Vaucouleurs law $r^{\frac{1}{n}}$ and the
galaxy luminosity.  The de Vaucouleurs $r^{\frac{1}{4}}$ law appears
to describe well only those galaxies with absolute magnitudes close to
$M_{V} \approx -19$ and where the region of the profile to be fitted
is restricted to approximately $0.2 < R / R_{\frac{1}{2}} < 1.5$
(e.g.\ Binggeli, Sandage \& Tarenghi 1984; Schombert 1986, 1987;
Burkert 1993; Coan et al.\ 1993; Graham et al.\ 1996). Early-type
galaxies brighter than this limit, including BCGs, are fit better with
an $n > 4$ profile, while galaxies fainter than this limit are fit
better by a profile which approaches an exponential disk ($n = 1$).
Using the median redshift of the galaxies in these fields ($\bar{z} =
0.76$; see Figure 3 of Paper I) and the spectral energy distribution
of a non-evolving elliptical, we can make a crude conversion to
absolute $V$ which implies that only $\sim 40\%$ of the galaxies in
these fields that we have visually classified as early-type galaxies
are brighter than $M_{V} = -19$. In addition, the MLE fittings cover a
significantly larger radial range than that specified above,
especially for the early-type galaxies where the half-light (or
effective) radii are small (see Figure~\ref{mdshlrg}). Both facts may
explain the observed relation between the best-fit model and the
visual classification of the early-type galaxies.

The automated classifications appear to agree better in the case of
late-type galaxies.  Here, 95\% of all galaxies visually classified as
spirals or irregulars contain a clear disk component (either D or D+B)
in their best-fit models. In the case of irregulars/peculiars, 81\% of
all galaxies with this classification are best-fit by a pure
disk. This result is further illustrated in Figure~\ref{mdsbtt} which
shows the distribution of bulge/(disk+bulge) luminosity ratios for
different morphological types. Early-type galaxies have
$\frac{B}{D+B}$ luminosity ratios which are spread between 0 (pure
disk) and 1 (pure bulge), whereas late-type galaxies clearly are
weighted much more heavily toward ratios of 0.

Figure~\ref{mdsab} shows the distributions of axial ratios for 44
ellipticals (E), 54 S0s, and 140 spirals (Sp) from the combined
cluster fields.  The axial ratios of the best-fit models are
plotted. In the cases where a galaxy is fit best by a disk+bulge
model, the axial ratio of the brightest component (disk or bulge) is
assumed. The distribution of axial ratios for elliptical galaxies show
a rapid decline from $\frac{b}{a} = 1$, while the S0 and spiral
distributions are significantly flatter.  The occasionally large
bin-to-bin variations are due to small number statistics, specifically
in the E and S0 samples. For example, the gap in the elliptical
distribution at $\frac{b}{a} = 0.85$ is a $\sim 2.5\sigma$
fluctuation.

The results of the distant sample are compared with the distribution
of axial ratios compiled by Sandage, Freeman \& Stokes (1970) using a
large sample of nearby field galaxies listed in the Reference
Catalogue of Bright Galaxies (de Vaucouleurs \& de Vaucouleurs
1964). They examined 168 E, 267 S0, and 254 Sp galaxies. (Irregular
galaxies were not included in their analysis.)  Their results indicate
that ellipticals have only moderate intrinsic flattening, while
ordinary spirals and S0s are intrinsically flatter (see Table 1 and
Figure 1 of Sandage, Freeman \& Stokes 1970). We have used the
$\chi^2$ test applicable for two binned data sets in order to confirm
that our axial ratio distributions are consistent with those of the
nearby galaxies. For each of the morphological populations (E, S0, and
Sp), the nearby and distant samples are consistent with a single
distribution function at a $> 99.9\%$ level. We have estimated the
isophotal limit of the Sandage, Freeman \& Stokes (1970) sample to be
only $\sim 0.5$ magnitudes brighter than ours (see Sect.\ 2.1). This
estimate is calculated by taking into account (1) the ${(1+z)}^{4}$
cosmological dimming, (2) the appropriate $k$-correction for each
galaxy type, and (3) the fact that the sky is $\sim 2$ magnitudes
fainter for space observations; in addition, because a surface
brightness limit is not specified in the Reference Catalogue, we have
assumed that this limit is roughly the same as that in the Second
Reference Catalog; that is, $\mu_{B} = 25~\rm{mag~per~arcsec^{2}}$ (de
Vaucouleurs, de Vaucouleurs \& Corwin 1976).

We have also compared our elliptical and S0 distributions to the
cluster sample of Andreon et al.\ (1996). This is a CCD survey of a
magnitude-limited sample of $\sim 100$ galaxies in the central region
of Coma.  Axial ratios measured at $\mu_{R} = 24.0~{\rm
mag~arcsec^{2}}$ are provided for each galaxy. Applying the
corrections listed above, their limit corresponds to an isophote which
is just $\sim 0.5$ magnitudes fainter than that used in our
analysis. We have used a two sample Kolmogorov-Smirnov (KS) test to
compare the axial ratios of our distant sample with their sample of 35
E and 35 S0 galaxies. We find that the two distributions are
consistent.  The probability that the distributions are drawn from the
same parent population is 10\% for the ellipticals and 20\% for the
S0s. Therefore, comparing surveys of roughly the same depth in surface
brightness, we find that the ellipticity distribution of our distant
galaxy sample is consistent with that of local field and cluster
galaxies.

Figure~\ref{mdshlrg} shows the distributions of the logarithm of the
half-light radius for different morphological types. There is clearly
a progression from early to late-type galaxies. Those galaxies
visually classified as spheroids are small and compact, with a median
half-light radius of 0.03 and 0.09 arcsec for the ellipticals and S0s,
respectively. The late-type galaxies are larger with typical
half-light radii of 0.37 and 0.17 arcsec for the spirals and
irregulars/peculiars, respectively. This trend results from the
correlation of galaxy morphology with central concentration and
surface brightness.  Late-type galaxies, on average, will be less
concentrated and less luminous than early-type galaxies (Morgan \&
Mayall 1957; Doi, Fukugita \& Okamura 1993; Abraham et al.\ 1996a;
Smail et al.\ 1997).  The observed range in half-light radii is
consistent with results from other fields observed with HST (Griffiths
et al.\ 1994; Casertano et al.\ 1995).

In Figure~\ref{arescomp}, we show the distribution of asymmetry
parameter (A) for different values of the visually-determined
disturbance index (D).  There appears to be a good correlation between
the two parameters. Our relation between the automated parameter $A$
and the visual parameter $D$ is very similar to the relation found by
the classifiers in Figure 2 of S97. This suggests that it is possible
to do a reasonable job at visually judging symmetry without being
significantly biased by an individual classifier. As previously
mentioned in S97, the disturbance index can do a better job at
estimating the individual galaxy asymmetry in the case where a galaxy
has a very close companion or lies on a background with a strong
gradient. In both cases, the simple, automated parameter used here
would overestimate the degree of asymmetry.

In summary, we have shown that there is a reasonable correlation
between the parameters of the visual and automated morphological
typing.  In the following analysis, we therefore use the visual
classifications of the brightest subsample of galaxies in each field
to quantify the cluster morphology.

\section{Cluster Galaxy Populations}

In this section, we present a detailed study of the properties of the
galaxy populations detected in the two cluster fields.  This includes
a general discussion of the colors and ages of all (field and cluster)
galaxies, in addition to a specific discussion of the morphological
composition of the cluster galaxy populations.

\subsection{Color--Magnitude Diagrams}

We have used the ground-based Keck imaging of the clusters to obtain
color information on our HST galaxies. Tables 4a,b lists the
corresponding $BVRI$ colors of the 209 and 205 galaxies in the
CL1604+4304 and CL0023+0423 fields, respectively, which have been
visually classified in the HST data (see Sect.\ 3.2). This information
includes the Keck photometric identification number of the
corresponding galaxy detected in the ground-based images and the
$BVRI$ aperture magnitudes (see Sect.\ 2.2). If the galaxy has a
measured redshift, its redshift and the characteristic age as
determined from the broad band AB values are listed (see also Sect.\
5.2). For more details on the galaxy photometry, spectra, and ages,
see Paper II.

The morphologically segregated $(B-R)$ versus $R$ and $(V-I)$ versus
$I$ color-magnitude (CM) diagrams of the two cluster fields,
CL0023+0423 and CL1604+4304, are presented in
Figures~\ref{00brmorph}--\ref{16vimorph}. Because of the better
resolution of the HST observations, an individual galaxy in the LRIS
image may be associated with more than one galaxy in the corresponding
HST image. This occurs in $\sim 10\%$ of the galaxies. Only the
brightest galaxies in each of these pairs are plotted in the CM
diagrams of Figures~\ref{00brmorph}--\ref{16vimorph}; these cases are
indicated in the notes to Table 4.  Galaxies with measured redshifts
are marked in these figures.  Foreground and background galaxies are
crossed out, while galaxies that are confirmed cluster members, as
determined from the velocity analysis of Paper II, are circled.  There
are 12 confirmed cluster members in each HST field of CL0023+0423 and
CL1604+4304. One of the cluster members in CL0023+0423 was not
detected in the MDS analysis (Keck \# 2166; see Sect.\ 5.2).

We examine the $(B-R)$ and $(V-I)$ colors because, in these bands, we
expect to see the largest change in color between a redshift of $z =
0$ and $z = 0.9$ for a non-evolving elliptical.  At redshifts
approaching $z = 0.9$, we expect $(B-R)_{\rm E/S0} \approx (V-I)_{\rm
E/S0} \sim 3 - 3.5$ for a non-evolving E or S0 galaxy (e.g.\ Fukugita
et al.\ 1995; Kinney et al.\ 1996).  In both fields, we see red,
elliptical cluster members that have $(B-R) \approx (V-I) \approx
2-3$. The observed colors are consistent with a passively evolving
galaxy population formed several Gyrs ago (see Sect.\ 5.2 and Paper
II).  We do not, however, see a very tight color-magnitude relation
for the early-type galaxies.  Such a relation is characteristic of
nearby and intermediate redshift clusters (e.g.\ Butcher \& Oemler
1984; Couch \& Newell 1984; Arag\'on-Salamanca et al.\ 1991; Luppino
et al.\ 1991; Dressler \& Gunn 1992; Dressler et al.\ 1994; Smail et
al.\ 1994; Stanford et al.\ 1995, 1997; Ellis et al.\ 1997).  In the
case of CL0023+0423, this appears simply to be the result of the small
number of ellipticals and S0s present in this cluster field. It is
obvious from the morphological classifications that this field is
comprised mainly of spirals and irregular/peculiar galaxies (see also
Sect.\ 5.4 and Figure~\ref{morphfrac}).  On the other hand,
CL1604+4304 contains a significantly larger fraction of early-type
galaxies; however, many of these galaxies are at or beyond the
completeness limit of the Keck photometric survey. At these faint
magnitudes, the photometric errors imply an error in the color which
is greater than $ 0.4^{\rm m}$, indicating that we would be
significantly washing out a CM relation whose scatter is typically $<
0.1^{\rm m}$ (e.g.\ Couch \& Newell 1984; Stanford et al.\ 1995, 1997;
Ellis et al.\ 1997).  Indeed, deep $HJK$ imaging reveals tight
optical--IR and IR--IR color-magnitude sequences (e.g.\
$\sigma_{(H-K)} \sim 0.08$) for the early-type galaxies in this
cluster (see Figure 2q of Stanford et al.\ 1997).

In Figure~\ref{brdist}, we show the distributions of $(B-R)$ colors
for various morphological types. This figure includes all of the
galaxies that are within the magnitude limits of the photometric
survey. There is a clear progression in color between early- and
late-type galaxies. As expected, the ellipticals and S0s are redder,
on average, than the spirals and irregulars/peculiars. The median
color of these distributions are $\{1.49~1.41~1.19~1.00\}$ for the
${\rm \{E~S0~Sp~Irr/Pec\}}$ bins, respectively. The distributions are
broad because we have included galaxies over a wide range in redshift.
The inset window in each panel shows the $(B-R)$ color distributions
of the confirmed cluster members. Because of the small numbers, we
have combined the data from both clusters, CL0023+0423 at $z = 0.84$
and CL1604+4304 at $z = 0.90$. Over this narrow range in redshift, we
expect the color difference for any morphological type to be less than
$\sim 0.2^{\rm m}$. Though the numbers are small, the median of the
elliptical color distribution appears to be redder than that of the
spirals.

\subsection{Redshifts and Color Ages of HST Galaxies}

In Figures~\ref{00hstz} and \ref{16hstz}, we show images of the
galaxies in the HST fields of CL0023+0423 and CL1604+4304,
respectively, that have measured redshifts. There are 41 and 29
galaxies, respectively, with measured redshifts in these cluster
fields.  The redshift is given in the upper left of each panel; the
two numbers at the bottom of each panel indicate the Keck photometric
identification number and the MDS object identification number,
respectively. In a few cases, some of these galaxies have not been
detected in the MDS automated object identification procedure (see
Sect.\ 3.1.1) and, therefore, no MDS identification number is listed.
In particular, in the CL0023+0423 field, Keck galaxy numbers 2792,
2166 (a cluster member), 2108, and 2003 are not detected. Each of
these galaxies was within $1^{''}$ of a WFPC2 CCD edge or on the
border between two CCDs. Because of the variable PSF in these regions,
these galaxies are excluded from the MDS analysis procedure (see
ROG). In the CL1604+4304 field, two very compact galaxies, Keck galaxy
numbers 1875 and 2515 (at $z = 0.4904$ and $0.4779$, respectively; see
Figure 14), are not included in the analysis as they were classified
as ``stellar'' in the MDS identification procedure.

As described in Paper II, we use the population synthesis models of
Bruzual \& Charlot (e.g.\ Bruzual 1983; Bruzual \& Charlot 1993;
Bruzual \& Charlot 1995) to determine the spectral ages of the
galaxies with measured redshifts.  Here, ``age'' refers to the time
(in Gyr) since the last period of {\it major} star formation. The free
parameters in these evolutionary models are the initial mass function
(IMF) and the star formation rate (SFR). We chose the traditional
Salpeter (1955) IMF with lower and upper mass limits of 0.1 and 125
$M_{\odot}$, respectively.  For the SFR, we have chosen an
exponentially decreasing SFR of $\Psi =
\tau^{-1}~e^{-{{t}\over{\tau}}}$ with a timescale $\tau = 0.6$ Gyr
(referred to as ``tau0.6'' in Paper II). The normalization implies
that it would take an infinite amount of time to convert all of the
galaxy's gas into stars (Bruzual 1983). The ages of the galaxies were
determined independently using their spectral energy distributions
(from the four AB magnitudes derived from the Keck photometric data)
and the equivalent widths of features in their spectra. According to
the convention of Paper II, the former is referred to as the ``color
age,'' while the latter is referred to as the ``spectral age.''  The
color ages appear to be more reliable, so we present these ages in the
following analysis. For details on the choice of the model and a
comparison between the two age determinations, see Sect.\ 4 of Paper
II.

In Figure~\ref{agez}, we plot the galaxy color age as determined from
the spectral energy distribution versus the galaxy redshift for each
cluster field. The morphology of the galaxy is indicated by different
symbols (the same symbols used in the color-magnitude diagram; see
Figures~\ref{00brmorph}--\ref{16vimorph}). Two points in the
CL0023+0423 field and four points in the CL1604+4304 panels have been
offset by $\pm 0.2$ Gyr to avoid overlapping any of the symbols. This
offset is roughly the average of the errors in the color ages (see
Table 4). For galaxies at redshifts greater than $z \sim 0.7$, it is
no longer possible to determine an accurate age from the galaxy colors
when their ages exceed $\sim 4$ Gyr. At this point, the relation used
to determine the age from the galaxy colors becomes flat, implying
that only lower limits can be placed on the galaxy age. This
degeneracy is discussed in Sect.\ 4 of Paper II (see also Figure 10 of
Paper II). In Figure~\ref{agez}, we have used arrows to indicate those
ages that are lower limits.

From Figure~\ref{agez}, we see that the majority of ages for the
late-type galaxies are $\simless 2$ Gyr. (These galaxies also have
strong \ion{O}{2} emission with a median equivalent width of 29 \AA;
see Paper II.) However, galaxies with ages $\simgreat 3$ Gyr are
predominately early-type galaxies, most notably in the CL1604+4304
field. This result can also be seen in Figure~\ref{agehist} which
shows the distribution of color ages as a function of morphological
class for the combined CL0023+0423 and CL1604+4304 fields. 83\% (35
out of 42) of the galaxies classified as late-type (spiral or
irregular/peculiar) have color ages $\simless 2$ Gyr. In contrast,
55\% (11 out of 20) of the early-type (elliptical and S0) galaxies
have color ages of greater than 2 Gyr, and approximately 73\% (8 out
of 11) of all galaxies with ages greater than 3 Gyr are classified as
early-type. The typical error in these percentages is $\sim 5 - 15\%$.
In addition, Figure~\ref{agez} shows that cluster galaxies are
typically older than field galaxies at similar redshifts. This is due
in part to the fact that there are more early-type galaxies in these
systems.

\subsection{Confirmed Cluster Members}

In this section, we present the relation between the galaxy morphology
and the spectral features in cluster galaxies. In particular, we
highlight the spectra of several cluster members in each of the two
fields. The galaxy spectra discussed below are shown in
Figures~\ref{00spcluster} and \ref{16spcluster} for clusters
CL0023+0423 and CL1604+4304, respectively.

\subsubsection{Early-Type Galaxies}

In CL0023+0423, there are four confirmed cluster members which are
classified as ellipticals (MDS ID \#17, 20, 45, 57), as well as one
compact galaxy classified as ``X'' (MDS ID \#155). Three of these five
(\#17, 20, 115) have typical elliptical spectra with \ion{Ca}{2} H and
K absorption, G-band absorption, and/or a 4000\AA\ break (e.g.\
Kennicutt 1992). These galaxies are red with $(V-I) \simgreat 2.48$
and have subsequent color ages of $\simgreat 2.7$ Gyr. However, the
other two galaxies (MDS ID \#45, 57) show star-formation features,
such as [\ion{O}{2}], and possibly H$\beta$ and [\ion{O}{3}]
emission. Consequently, these galaxies are bluer than the other
ellipticals with $(V-I) = 2.23$ and $1.38$ for MDS ID \#45 and 57,
respectively. Each of these galaxies also has a noticable asymmetric
disk (see Figure~\ref{00hstz}) and has been classified as possible
mergers (see Table 2a).  Because of the fairly strong, narrow emission
lines, their spectra appear to be more typical of blue, compact
galaxies found at redshifts $z \sim 0.10 - 0.66$ in previous HST
observations (Koo et al.\ 1994, 1995). The two galaxies have a compact
shape, relatively blue color [{($B-V)_{\rm o} \sim 1.38$ and $0.55$,
respectively], and a relatively high luminosity ($M_{B} \approx
-19.97$ and $-19.52 + 5~{\rm log}~h$, respectively). All of these
properties are consistent with this class of galaxies.  Because of
such characteristics, \ion{H}{2} galaxies have been suggested to be
the likely local counterparts of blue compact galaxies (Terlevich
1987; Terlevich et al.\ 1991; Koo et al.\ 1994, 1995). The blue colors
and strong emission lines suggest a recent, strong burst of star
formation.  Koo et al.\ (1995) suggest that such an event would be
followed by several magnitudes of fading, resulting in a galaxy with a
surface brightness and velocity width which are typical of nearby
low-luminosity spheroids (see Binggeli, Sandage \& Tammann 1985;
Kormendy \& Bender 1995). Therefore, blue compact galaxies can be the
progenitors of contemporary spheriodal galaxies. The wide range in
redshift over which these galaxies are found implies that major star
formation episodes have occurred in some spheriods over many Gyrs.

In CL1604+4304, there are three cluster members that are visually
classified as ellipticals (MDS ID \# 9, 13, 50) and one that is
classified as an S0 (MDS ID \# 65). All of the ellipticals are red
with $(V-I) \simgreat 2.76$; have typical absorption spectra with
\ion{Ca}{2} H and K absorption, G-band absorption, and a 4000 \AA\
break; and color ages of $\simgreat 3.5$ Gyr (see Sect.\ 5.2). In
every way, these galaxies represent the quiescent early-type cluster
population whose behavior is consistent with passive evolution of an
old stellar population formed in a exponentially decaying burst of
star formation at redshifts of $z > 2$ (Arag\'on-Salamanca et al.\
1993; Rakos \& Schombert 1995; Dickinson 1995; Steidel et al.\ 1996;
Ellis et al.\ 1997; Stanford, Eisenhardt \& Dickinson 1997; Paper
II). The S0 galaxy (MDS ID \# 65), on the other hand, is bluer with
$(V-I) = 1.96$, a color age of 1.7 Gyr, and spectrum that shows
[\ion{O}{2}] and H$\beta$ emission. Such star-forming features in
spectra of S0 galaxies at intermediate and high redshift are not
uncommon (Trager 1997; Poggianti 1997).  Indeed, if S0 galaxies form
out of spirals through ram pressure stripping (e.g.\ Larson, Tinsley
\& Caldwell 1980), galaxy-galaxy interactions (e.g.\ Moore et al.\
1996), or mergers, such spectroscopic evidence of recent or current
episodes of star formation should be common in this population. In
this case, the morphology, including an asymmetric lens and a close
companion (see Figure~\ref{16hstz}), may also indicate the occurrence
of these environmental effects.

\subsubsection{Spiral, Irregular and Peculiar Galaxies}

In each cluster, there is one galaxy classified as a late-type which
is unusually red (see Figures~\ref{00brmorph}--\ref{16vimorph}).  At
the redshifts of these clusters, we would expect that a non-evolving
spiral later than Sb to have colors of $(B-R) \approx (V-I) \simless
2$.  In CL0023+0423, we find a cluster member (MDS ID \#82) classified
as an Irr/Pec which has $(B-R) = 2.85$ and $(V-I) = 2.70$. This galaxy
is situated on the border between the PC and a WFPC CCD. Consequently,
only the diffuse extension of the galaxy's disk was detected in the
automated MDS procedure and, therefore, visually classified as ``P''
for peculiar. Though it is possible that this diffuse emission is
another galaxy, the system as a whole resembles a peculiar Sa galaxy
(see Figure~\ref{00hstz}). For a non-evolving Sa galaxy, the typical
colors at these redshifts would be $(B-R) \sim 3.0$ and $(V-I) \sim
2.6$ (Kinney et al.\ 1996). Therefore, this cluster galaxy has a color
that is close to that of a non-evolving early spiral galaxy,
indicating little color evolution. This is also consistent with the
galaxy spectrum which shows \ion{Ca}{2} H and K and a 4000 \AA\ break
and with the $K^{'}$ survey at the KPNO 4m (see Paper I) which gives
$(V-K^{'}) \sim 6$ for this galaxy (Lubin, Oke \& Postman 1998).

Similarly, in the cluster CL1604+4304, there is a spiral galaxy (MDS
ID \#7) that is quite red, $(B-R) = 2.32$ and $(V-I) = 3.37$, and
which has an estimated age of 3.8 Gyr.  This galaxy has been
classified as an Sa.  In addition, there is clear evidence of a tidal
interaction with a close companion (see Figure~\ref{16hstz}),
indicating that this galaxy has likely suffered an abrupt change in
its star-formation rate. A burst and/or truncation of star formation
on a short timescale would make a representative galaxy become
temporarily blue with emission lines and decay through a
``post-starburst'' spectral phase to become an optically-red
system. Such signatures are often seen in detailed spectroscopic and
photometric studies of intermediate redshift clusters (e.g.\
Arag\'on-Salamanca et al.\ 1991; Couch et al.\ 1994). This galaxy has
a spectrum which resembles an ``E+A'' galaxy. An ``E+A'' spectrum is
dominated by a young stellar population but lacks the strong emission
lines characteristic of on-going star formation (Dressler \& Gunn
1983; Gunn \& Dressler 1988; Zabludoff et al.\ 1996). Such a spectrum,
which shows strong Balmer absorption lines, implies that the galaxy
has experienced a brief starburst within the last Gyr (Dressler \&
Gunn 1983; Couch \& Sharples 1987). In this case, the starburst was
apparently the result of a galaxy-galaxy interaction since there is
clear evidence of a tidal tail.

There are also several normal late-type galaxies in both of these
clusters with spectra characterized by [\ion{O}{2}] emission and ages
less than $\sim 2$ Gyr (see Sect.\ 5.2).  In addition, we see some
disturbed late-types and peculiar galaxies. In CL0023+0423, MDS ID \#
37 is a disturbed Sc which contains a double nucleus. Its colors are
very blue with $(B-R) = 0.60$ and $(V-I) = 1.50$, implying a color age
of 0.8 Gyr. Its spectrum contains extremely strong [\ion{O}{2}],
H$\beta$, and [\ion{O}{3}] emission with equivalent widths of 82.7,
12.4, and 98.6 \AA\, respectively (see Tables 2 and 3 of Paper
II). The strong emission and the double nucleus suggest a recent
merger. In CL1604+4304, there is another such example, MDS ID \#
25. This galaxy contains two compact, high surface-brightness
nuclei. It has intermediate colors of $(B-R) = 1.35$ and $(V-I) =
1.99$ and a color age of 1.7 Gyr. Its spectrum contains strong
[\ion{O}{2}] emission with an equivalent width of 29 \AA, indicating
active star formation, plus \ion{Ca}{2} K absorption.  This system may
be an elliptical merger.

In almost all cases, strong \ion{O}{2} emission is associated with
late-type galaxies. Of all confirmed cluster members in the
CL0023+0423 and CL1604+4304 systems, 78\% of all galaxies with
\ion{O}{2} equivalent widths of greater than 15 \AA\ are classified as
spiral or irregular/peculiar galaxies. The remaining 22\% are
classified as ellipticals, though their spectral and photometric
properties indicate that they are more similar to blue compact
galaxies (see Sect.\ 5.3.1).

Figures~\ref{00spcluster} and \ref{16spcluster} show the galaxy
spectra discussed above for clusters CL0023+0423 and CL1604+4304,
respectively.  These spectra reveal some of the difficulties
associated with faint object spectroscopy. For example, in some cases,
poor sky subtraction leaves obvious residual sky lines at 5577 \AA,
5891 \AA, and 6300,6363 \AA\ in the blue end of the spectrum and at
$\simgreat 8000$ \AA\ in the red end of the spectrum (see a sample sky
spectrum in Figure 3 of Paper II). In addition, identified lines in
the near-infrared that may not seem convincing due to the large number
of residual sky lines in this region are actually obvious in the
two-dimensional spectrum (see Sect.\ 4.2.1 of Paper I for the details
of the line identification and redshift determination). The average AB
magnitude error in these spectra is $\sim 0.14^{\rm m}$ at 7500 \AA.

As discussed above, the individual spectra show that, in general, the
morphologies of the cluster galaxies appear to be consistent with the
galaxy types and features (e.g.\ interactions or mergers) that one
would predict based on their spectral features alone.

\subsection{Statistical Distribution of Cluster Galaxy Morphologies}

One of the primary goals of this investigation is to study the overall
morphological composition of each cluster. We cannot, however, examine
all the cluster members on an individual basis since our direct
redshift measurements are limited in the HST field-of-view. Therefore,
in this section, we examine the {\it background-subtracted} morphology
distribution in each cluster. In order to determine the contribution
of the background field galaxies, we use the morphologically
classified galaxies from the Medium Deep Survey (MDS; Griffiths et
al.\ 1994) and the Hubble Deep Field (HDF; Williams et al.\ 1996).

We use the MDS + HDF galaxy number counts in the F814W ($I_{814}$)
filter presented in Abraham et al.\ (1996a,b). There is a discrepancy
in the total MDS number counts plotted in Figure 7 of Abraham et al.\
(1996a) and those replotted with the total HDF number counts in Figure
6 of Abraham et al.\ (1996b). Therefore, to avoid any uncertainty, we
have used directly the tables of galaxy magnitudes and morphological
classifications presented in Table 1 of Abraham et al.\ (1996a) for
the MDS counts and Table 1 of van den Bergh et al.\ (1996) for the HDF
counts. The total effective area for the MDS and HDF survey are
$82.9~{\rm {arcmin}^2}$ and $3.95~{\rm {arcmin}^2}$, respectively
(R.G. Abraham, private communication). We have used the galaxy
classification of S.\ van den Bergh to split the galaxy counts of the
combined MDS and HDF fields into early-types (E/S0), spirals (Sp), and
irregulars/peculiars (Irr/Pec). Their Irr/Pec bin, like our bin of the
same name, includes mergers, interactions, and those galaxies which
are simply unclassifiable according to the standard Hubble scheme.
Even though we have used van den Bergh's galaxy classifications, these
classifications and those by R.S.\ Ellis and the automated ``machine''
analysis all give consistent morphologically-segregated number counts
(Abraham et al.\ 1996a,b). In order to combine the two field datasets,
the resulting number counts of the four divisions (total, E/S0, Sp and
Irr/Pec) are modeled as either a power-law or a power-law plus an
exponential cut-off, depending on the shape of the distribution. The
best-fit analytic functions are integrated over the appropriate
magnitude range in order to determine the expected number of
background galaxies per unit area for our two cluster fields. This
method means that we can accurately model the galaxy counts as a
function of magnitude in each morphological bin even though the
absolute morphological fractions have changed from the shallower MDS
survey to the deeper HDF survey (see Abraham et al.\ 1996a,b; van den
Bergh et al.\ 1996).

For the cluster field CL1604+4304, we can directly use the background
counts determined in the above analysis as the HST observations for
this field were also taken in the F814W filter. For example, at
$I_{814} = 24.3$, our total magnitude limit for the visual
classification (see Sect.\ 2.1), the foreground/background
contamination is $\Sigma_{814} = 36.6~{\rm
galaxies~{arcmin}^{2}}$. The breakdown in morphology is roughly 22\%
E/S0, 44\% Sp, and 34\% Irr/Pec, with typical errors in each class of
$\sim 5-8\%$. We also obtain the appropriate split between ellipticals
and S0s in our E/S0 bin by examining the combined MDS+HDF data
(Abraham et al.\ 1996a; van den Bergh et al.\ 1996). The field ratio
of E:S0 ranges from approximately 1:1 at the bright end ($I_{814} <
22$) to approximately 3:1 at the faint end ($22 < I_{814} < 25$).  At
our magnitude limit, this ratio is roughly 2:1. Therefore, we have
adopted this ratio for the following morphological analysis.

Because we are using field data whose galaxies have been analyzed and
classified by other observers, we have examined several other sources
in order to ensure that our background galaxy estimates are
reasonable. Firstly, we have performed a similar analysis on data
kindly provided by S.P. Driver of a deep MDS field with independent
morphological classifications (Casertano et al.\ 1995; Driver,
Windhorst \& Griffiths 1995; Driver et al.\ 1995). Their galaxy number
counts imply a field contamination of $\Sigma_{814} = 35.2~{\rm
galaxies~{arcmin}^{2}}$ and a morphological mix of 17\% E/S0, 38\%
Sabc, and 45\% Sd/Irr.  The total and the early-type number density of
galaxies are consistent, within the errors assuming Poisson
statistics, with the analysis of the Abraham et al.\ data.  The data
of Driver et al.\ are binned slightly differently with late-type Sd
galaxies being included in the irregular bin. If this segregation had
been made in the Abraham et al.\ data, the percentages would again be
consistent within the errors. Finally, we have examined the deep $I$
band Keck counts of Smail et al.\ (1995). These data reach $I = 25.5$,
and the passband is very similar to $I_{814}$, with $I_{814} = I +
0.05$. If we integrate the best-fit power-law function over the same
magnitude range, we find $\Sigma_{814} = 37.5~{\rm
galaxies~{arcmin}^{2}}$, consistent with the above results.

The analysis of the cluster field CL0023+0423 is less
straightforward. The HST observations of this field were taken in the
F702W ($R_{702}$) filter; therefore, we need to convert the background
galaxy counts in the F814W to the F702W filter. In order to accomplish
this, we use the redshift distribution of galaxies in the HDF to
determine the median redshift of the field population. The redshift
data are being compiled by groups at Caltech and the University of
Hawaii Institute for Astronomy (Cohen et al.\ 1996). The median redshift
of the HDF is $z = 0.53$ ($\sigma = 0.22$). We use the non-evolving
spectral energy distributions (SEDs) of Coleman, Wu \& Weedman (1980)
to calculate the $(R_{702} - I_{814})$ colors of the relevant
morphological types at this redshift. We find ($R_{702} - I_{814}) =
\{0.53~0.29~0.24\}$ for a $\rm {\{E~Sbc~Sdm\}}$ galaxies,
respectively. These colors are used to make an {\it average}
conversion from the $I_{814}$ number counts. At $R_{702} = 24.7$, our
total magnitude limit for the visual classification (see Sect.\ 2.1),
the background contamination is $\Sigma_{702} = 37.2~{\rm
galaxies~{arcmin}^{2}}$, with a morphological mix of roughly 22\%
E/S0, 44\% Sp, and 34\% Irr/Pec. In addition, we have confirmed these
results by converting the deep $R$ band Keck counts (Smail et al.\
1995) to $R_{702}$ by using the Holtzman et al.\ (1995a) conversion
for a galaxy with a characteristic color of $(V-R) = 0.6$ (see Figure
3 of Smail et al.\ 1995). At $R_{702} = 24.7$, we find $\Sigma_{702} =
37.7~{\rm galaxies~{arcmin}^{2}}$, consistent with the above results.

The field distributions determined are subtracted from the cluster
morphological distributions. In Figure~\ref{morphfrac}, we show the
field-subtracted distributions of galaxy morphology for galaxies
brighter than $M_{V} = -19.0 + 5~{\rm log}~h$ in CL0023+0423 and
CL1604+4312. We have transferred to the cluster $V$ rest frame by
using the observed $R_{702}$ or $I_{814}$ total magnitudes (see Lubin
1996 for details on such a transformation).  The $k$-correction at the
cluster redshift and the rest-frame ${(V - R_{702})}_{\rm o}$ or ${(V
- I_{814})}_{\rm o}$ color are computed by convolving the non-evolving
SEDs of Coleman, Wu \& Weedman (1980) for each morphological type with
the system filter bandpasses (see also Frei \& Gunn 1994; Fukugita et
al.\ 1995; Kinney et al.\ 1996).  For both clusters and all galaxy
types, this absolute magnitude limit is equal to or less than the
magnitude limit down to which we have visually classified
galaxies. Galaxies detected in the PC have not been included in this
analysis.

The most obvious result in Figure~\ref{morphfrac} is the strikingly
different distribution of morphologies between the two cluster
fields. CL0023+0423 is dominated by spiral galaxies, while CL1604+4304
is composed mainly of early-type galaxies.  The general properties of
these field-corrected distributions are quite robust. We have
confirmed this by, firstly, varying the magnitude and morphological
mix of the background distribution based on the variations in the
values of the field data discussed above. This includes both the
uncertainty in the E/S0 split and the possible range in the conversion
from the $I_{814}$ to $R_{702}$ field counts. Secondly, we have
examined the effect of the $k$-correction on these distributions by
using an {\it evolving} spectral energy distribution, rather than the
assumed no-evolution models (see above).  The evolving SED is derived
from the adopted $\tau = 0.6~{\rm Gyr}$ Bruzual \& Charlot model (see
Sect.\ 5.2).  For each galaxy type, we have determined the appropriate
age of the model from the actual observational results (see
Figure~\ref{agez}).  We find that variations of both kinds do make a
small quantitative difference in the absolute numbers, especially in
those bins which contain relatively few galaxies. For example, the
elliptical bin in the cluster CL0023+0423 contains $-4 \pm 5$
galaxies. We know, however, that there are at least four confirmed
cluster members which have been classified as an elliptical (see Table
2 and Sect.\ 5.3). Though we are consistent with Poisson statistics at
a 1.6-$\sigma$ level, variations of this order can also be produced by
the uncertainty in the adopted field distribution and the
$k$-correction. However, the qualitative behavior of the two
distributions remains the same even if we adopt the largest deviations
in these two quantities.

CL0023+0423 has a galaxy population that is more typical of the
field. The numbers from the statistical analysis are consistent with
100\% of the galaxies in this cluster being normal spirals; however,
we know that, in the HST field-of-view, there are at least three (out
of 12) confirmed cluster members which have both the morphological and
spectral characteristics of an elliptical (see Sect.\ 5.3.1). If we
assume that the errors in the total and field galaxy counts are due to
Poisson fluctuations, a spiral fraction as low as 55\% (1-$\sigma$)
could be possible, completely consistent with the spectral results.

CL1604+4304, in contrast, has a morphological composition which is
more characteristic of a normal, present-day rich cluster. Early-type
galaxies comprise $76^{+24}_{-27}\%$ of all galaxies in the central
$\sim 0.5~h^{-1}~{\rm Mpc}$ of this cluster. The proportion of S0
galaxies and ellipticals is 48\% and 28\%, respectively. This implies
a ratio S0/E of $1.7 \pm 0.9$ which is consistent with galaxy
populations found in local clusters. Dressler (1980a) found an average
value of S0/E $\sim 2$ for a sample of 11 clusters at $0.035 < z <
0.044$. The survey included all galaxies brighter than $M_{V} = -20.4$
with $h = 0.5$ and $q_{0} = 0.5$.  We note that the exact S0/E ratio
is far from certain because of the difficulty in distinguishing
between elliptical and S0 galaxies at this redshift (see Sect.\ 3.2
and Discussion).

The fraction of S0 galaxies in CL1604+4304 is higher than those found
in recent studies of rich, intermediate-redshift clusters at $z =
0.37-0.56$. Classifiers of these cluster galaxies find elliptical
fractions that are comparable to present-day clusters; however, the S0
fractions are smaller than nearby cluster populations by a factor
$\sim 2 - 3$. The ratios of S0/E for these intermediate-redshift
clusters are typically less than 0.5 (D97). Of course, the results
from our survey and that of D97 depend strongly on the adopted ratio
of E:S0 in the field population. The survey of D97 reaches a brighter
limiting magnitude of $I_{814} = 23$.  For their morphological
analysis, the authors have chosen a lower field ratio of ${\rm E:S0} =
1:1$.  If we had adopted a similar mix, we would find a ratio of S0/E
of $1.1 \pm 0.5$, still higher than that found by D97. However,
classifications of field galaxies at $I_{814} > 22$ seem to indicate
field ratios of E:S0 which are greater than 2:1 (van den Bergh et al.\
1996).

The morphological results support the conclusions of the dynamical
analysis presented in Paper II. This study also indicates that the two
clusters are very different in nature (see Paper II).  CL0023+0423
consists of two substructures with mean redshifts of 0.8274 and 0.8453
and velocity dispersions of $158^{+42}_{-33}$ and
$415^{+102}_{-63}~{\rm km~s^{-1}}$, respectively. The two systems are
separated in velocity by approximately 2922 ${\rm km~s^{-1}}$ and are
separated in the plane of the sky by $\sim 229$ kpc.  The virial and
projected mass estimates are $1.0^{+0.5}_{-0.4} \times 10^{13}$ and
$3.6 \pm 0.5 \times 10^{13}$ $h^{-1}~{\rm M_{\odot}}$, respectively,
for the low dispersion substructure and $2.6^{+1.3}_{-0.8} \times
10^{14}$ and $4.2 \pm 0.7 \times 10^{14}$ $h^{-1}~{\rm M_{\odot}}$,
respectively, for the high dispersion substructure (Paper II). The
velocity dispersions, masses, and morphological composition indicate
that these systems are similar to local groups of galaxies (e.g.\
Ramella, Geller \& Huchra 1989; Zabludoff \& Mulchaey 1997 and
references therein). Though this may be a chance projection, the
dynamical and morphological evidence may imply that we are seeing the
merger of two spiral-dominated galaxy groups (see Lubin, Postman \&
Oke 1998a).

On the other hand, CL1604+4304 has a mean redshift of 0.8967 and a
velocity dispersion of ${\rm 1226^{+245}_{-154}~km~s^{-1}}$ (Paper
II). The velocity histogram is consistent with a Gaussian
distribution, implying that the cluster is already well-formed and
relaxed. The virial and projected mass estimators give
$7.8^{+3.2}_{-2.1} \times 10^{14}$ and $2.5 \pm 0.2 \times 10^{15}$
$h^{-1}~{\rm M_{\odot}}$, respectively.  Furthermore, this cluster was
detected in X-rays by ROSAT with a bolometric X-ray luminosity of $L_x
\sim 2 \times 10^{44}~h^{-2}~{\rm erg~s^{-1}}$ (Castander et al.\
1994). The X-ray--optical properties of this cluster are consistent
with the local $L_{x} - \sigma$ relation (Mushotzky \& Scharf
1997). The velocity dispersion, cluster mass, X-ray luminosity, and
morphological composition are all consistent with Abell richness class
2 and 3 clusters (Dressler 1980a; Bahcall 1981; Struble \& Rood 1991;
Mushotzky \& Scharf 1997).

\section{Discussion}

The most intriguing result of this study is the striking difference in
the statistical morphological distributions of CL0023+0423 and
CL1604+4304. These distributions indicate that CL0023+0423 has a
galaxy population which is more typical of the field. The numbers are
consistent with 100\% of the galaxies in this cluster being normal
spirals (though the spectral results indicate that there are at least
a few ellipticals in this system; see below). The velocity analysis
reveals that this cluster actually consists of two smaller systems
with individual dispersions of 158 and $415~{\rm km~s^{-1}}$ and
separated by $\sim 2922~{\rm km~s^{-1}}$. The velocity dispersion and
dynamical mass of each system are more typical of galaxy groups and
poor clusters. The dynamical analysis of this two-body system is
consistent with both a bound and unbound solution (Lubin, Postman \&
Oke 1998a); therefore, we cannot say for certain that these two
systems are in the process of merging.  However, high-redshift groups
of galaxies such as these are likely to be the building blocks of
intermediate-redshift clusters, most certainly if theories of
hierarchial clustering are valid.  It seems reasonable that, if such
systems were to combine to form a cluster themselves, rather than
simply be accreted by a larger system, they may be the progenitors of
open clusters at intermediate redshift. These clusters are irregular,
loose, and presumably dynamically young. Studies of this class of
clusters at $z = 0.31- 0.54$ indicate that they have elliptical
fractions between $27 - 47\%$ and total early-type (E + S0) fractions
between $45 - 80\%$ (D97; Stanford, Eisenhardt \& Dickinson 1997;
Andreon, Davoust \& Heim 1997; Couch et al.\ 1988).

The CL0023+0423 system does contain three (out of twelve) galaxies
which are morphologically classified as either an elliptical or a
compact galaxy and which have photometric and spectral properties
indicating that it formed at a redshift of $z > 3$. The overall
statistical distribution of morphologies in CL0023+0423 implies that
early-type galaxies may comprise only $5^{+35}_{-5}\%$ of the total
population, consistent with the 25\% from the confirmed group members
in the HST field-of-view. If we try to improve the statistics by
examining all confirmed cluster members in the larger LRIS
field-of-view, we still find that only 17\% (4 out of 24) of the
galaxies have a typical elliptical-like, absorption spectrum.
Two-thirds of all confirmed members have very strong \ion{O}{2}
emission (equivalent widths of typically much greater than 10 \AA; see
Paper II).  Given the correlation between active star formation and
galaxy morphology (see Sect.\ 5.3), this implies an early-type
population of 33\% or less, consistent with the numbers discussed
above. If these groups do combine to make irregular clusters observed
at intermediate redshifts, a non-negligible fraction of early-type
galaxies may be forming between redshifts of $z \sim 0.9$ and $z \sim
0.5$. Therefore, we would expect to see a significant fraction of
early-type galaxies in open clusters that have spectral features
characteristic of star formation activity within the last $\sim 1$
Gyr. Preliminary spectral studies of a sample of 10
intermediate-redshift clusters, including both open and compact
clusters, indicate that the bulk of the early-type population has
passive spectra with no signs of current or recent star formation;
however, there is a non-negligible fraction which show post-starburst
spectral features (Poggianti 1997).

In addition, the modest early-type fractions in both galaxy groups of
the CL0023+0423 system imply that the strong correlation between
velocity dispersion and early-type fraction observed in nearby groups
of galaxies (Zabludoff \& Mulchaey 1997) does not exist at high
redshift.  Based on the local relation, we would expect an early-type
fraction of $f_e \sim 0.10$ for our low dispersion system of $\sigma
\sim 158~{\rm km~s^{-1}}$ and $f_e \sim 0.55$ for our high dispersion
system of $\sigma \sim 415~{\rm km~s^{-1}}$ (see Figure 7 of Zabludoff
\& Mulchaey 1997); however, the spectroscopic results suggest that the
observed fractions may be as low as $\sim 0.29$ (2 out of 7) and $\sim
0.12$ (2 out of 17) for the low and high dispersion systems,
respectively.  We note that the a priori probability of finding only
two early-types out of 17 galaxies when the probability for success
is 55\% is an unlikely $\sim 0.03\%$.  If the two groups of
CL0023+0423 are typical of galaxy groups at high redshift, and they
are the progenitors of local groups, there appears to be a progression
in the group morphological composition between redshifts of $z \sim
0.8$ and the present epoch. At high redshift, there is no apparent
correlation between velocity dispersion and early-type fraction;
galaxy groups of varying mass all appear to have relatively low
fractions of spheroids.  This result may indicate that there is
continual elliptical and S0 formation at redshifts of $z < 1$ and that
these galaxies will form only in relatively massive regions of the
universe.

CL1604+4304 has a morphological composition which is characteristic of
a normal, present-day rich cluster. From the statistical distribution
of morphologies, we find that early-type galaxies comprise
$76^{+24}_{-27}\%$ of all galaxies in the central $\sim
0.5~h^{-1}~{\rm Mpc}$ of this cluster. The dynamical analysis
indicates that this system is already well-formed and relaxed.  The
velocity dispersion, cluster mass and X-ray luminosity are consistent
with an Abell richness class 2 or 3 cluster (Dressler 1980a; Bahcall
1981; Zabludoff, Huchra \& Geller 1990; Struble \& Rood 1991;
Castander et al.\ 1994; Mushotzky \& Scharf 1997).  Of the early-type
cluster population, the ratio of S0 galaxies to ellipticals is $1.7
\pm 0.9$, consistent with galaxy populations found in local clusters
(Dressler 1980a).  The fraction of S0 galaxies is higher than those
found in recent studies of rich, intermediate-redshift clusters at $z
= 0.37-0.56$. The studies of D97 indicate elliptical fractions which
are comparable to present-day clusters; however, the S0 fractions are
smaller than nearby cluster populations by a factor $\sim 2 - 3$. The
ratios of S0/E for these intermediate-redshift clusters are typically
less than 0.5 (D97). These findings imply that the elliptical
population is already in place by $z \sim 0.5$; in contrast, a large
fraction of the S0 galaxies are still forming between redshifts of $z
\sim 0.5$ and $z = 0$, presumably out of the excess of late-type
galaxies.  However, these morphological fractions are far from
certain.  For example, Stanford et al. (1997) have independently
studied the HST data for 19 galaxy clusters with redshifts between $z
\sim 0.3 - 0.9$. Their sample includes CL1604+4304 and eight of the
ten intermediate-redshift clusters studied by D97. They find that
these clusters contain early-type (E+S0) fractions that are consistent
with the fraction that we find in CL1604+4304; in addition, the
fraction remains roughly constant over their entire redshift range
(Stanford et al.\ 1997). This result is inconsistent with that of D97
as it implies that there is no decline in the frequency of early-type
cluster galaxies with redshift (see also Andreon, Davoust \& Heim
1997; Andreon 1998). This decline would be expected if, as suggested
by D97, there exists an elliptical population which remains
effectively constant and a S0 population which is increasingly
depleted with redshift because it has not yet formed from the
late-type galaxies.

If the morphological fractions of D97 are indeed correct, it would
imply that CL1604+4304 is not the progenitor of the
intermediate-redshift clusters that they have studied; it may,
however, be a cluster which formed at a much earlier epoch with enough
time to create a comparable fraction of S0 galaxies. If a similar
morphological transformation is taking place in this cluster, we would
expect a population of ellipticals that have passive spectra and a
population of S0 galaxies that exhibit spectral indications of recent
and/or current star formation. Though the number of confirmed cluster
members is small, there are three ellipticals and one S0 galaxy in
this cluster. The three ellipticals all have typical red, passive
spectra and color ages implying the last major period of star
formation was greater than 3.5 Gyr ago (a formation epoch of $z >
5$). The S0 galaxy, on the other hand, is bluer, has a color age of
1.7 Gyr, and has a spectrum that shows [\ion{O}{2}] and H$\beta$
emission. Though the numbers are too small to say anything with
certainty, the morphology--spectral properties are consistent with a
cluster that has an old population of ellipticals and an evolving
population of S0 galaxies. We can try to examine this behavior over
the larger LRIS field-of-view by sorting all of the cluster members
according to their spectral characteristics. If we make the broad, and
not completely valid, assumption that (1) all galaxies with a typical
K star absorption spectrum are ellipticals, (2) all galaxies with a
spectrum that contains K star plus emission features are S0 galaxies,
and (3) all galaxies with a pure emission-line spectrum are spiral and
irregular/peculiar galaxies, we find the percentages of elliptical and
S0 galaxies are 32\% (7 out of 22) and 36\% (8 out of 22),
respectively. Though this is not formally correct as the spectral
properties of a particular morphological class can vary from galaxy to
galaxy, these numbers are consistent with the statistical distribution
of galaxy morphologies in this cluster.

The morphology--density relation is observed in both open and compact
clusters in the local universe (see Introduction); however, this is
apparently not the case at intermediate redshifts.  D97 find that the
morphology--density relation is qualitatively similar in compact
clusters at intermediate redshifts, but completely absent in the open
clusters at a similar epoch. The authors suggest that this result
implies that morphological segregation occurs hierarchically over
time. The richer, denser clusters, which obviously form at an early
epoch, are affected first. The smaller, less dense systems, which are
younger dynamically, form later in time and, therefore, the
segregation has not proceeded as far. That is, groups which make up
the irregular, open clusters at intermediate redshifts have not
undergone significant morphological segregation to establish a
morphology--density relation; however, by the present epoch, the
groups which make up the local open clusters (such as the Virgo and
Hercules cluster) would have had sufficient time to establish such a
correlation. If this hypothesis is correct, clusters at high redshift
should show little or no morphological segregation. In light of this,
we are in process of studying the morphology--density relation for our
sample of clusters at $z > 0.7$ (Lubin, Postman \& Oke 1998b).

\section{Conclusions}

As part of an observational program to study distant clusters of
galaxies, we have examined the morphological properties of the
galaxies in two cluster fields, CL0023+0423 at $z = 0.84$ and
CL1604+4304 at $z = 0.90$, using high-resolution HST images. The
morphology of the individual galaxies have been studied by two
methods; 1) a quantitative description of the structural properties of
$\sim 600$ galaxies per cluster field is provided by the Medium Deep
Survey automated data reduction and ``bulge+disk'' object
classification software; 2) the brightest subsample of $\sim 200$
galaxies per cluster field are assigned a more detailed morphological
description through a visual classification based on the revised
Hubble scheme. A comparison between the two techniques shows that
there is a reasonable correlation between the parameters of the
automated and visual classifications (see also Lahav et al.\ 1995). To
investigate the morphological composition of the two galaxy clusters,
we have used the visual classifications of the brightest subsample of
galaxies in each field.  Our main conclusions are summarized below.

\newcounter{discnt}
 
\begin{list}
{\arabic{discnt}.}  {\usecounter{discnt}}

\item The color-magnitude diagrams and the color histograms of all
(field + cluster) galaxies in the two cluster field show a clear
progression in color between early- and late-type galaxies. As
expected, the elliptical and S0 galaxies are redder, on average, than
the spirals and irregulars. This trend is also apparent in the color
ages which represent the time since the last period of major star
formation. 83\% of the galaxies classified as late-type (spiral or
irregular/peculiar) have color ages less than 2 Gyr. In contrast, 55\%
of the galaxies classified as early-type have color ages of greater
than 2 Gyr, and 73\% of all galaxies with ages greater than 3 Gyr are
classified as early-type. In addition, cluster galaxies are typically
older than field galaxies at similar redshifts. This is due in large
part to the fact that there are more early-type galaxies in these
systems.

\item We see a distinct correlation between the galaxy morphology and
the corresponding spectral features. We have specifically examined
this relation for those galaxies which are confirmed cluster
members. The majority of galaxies that are visually classified as
ellipticals show spectra which are typical of nearby, red elliptical
galaxies. However, some of the galaxies visually classified as
ellipticals turn out to be blue compact galaxies with spectra
characterized by fairly strong, narrow emission lines. Normal
late-type galaxies typically have spectra with blue colors and
[\ion{O}{2}] emission, while the presence of strong star-formation
features, such as extremely high equivalent width [\ion{O}{2}], ${\rm
H\beta}$, and/or [\ion{O}{3}] emission, is always accompanied by
peculiar morphologies which suggest recent mergers or interactions.

\item Despite being at very similar redshifts, the two cluster systems
contain very different galaxy populations as indicated by their
background-subtracted morphological distributions.  We have examined
all galaxies brighter than $M_{V} = -19.0 + 5~{\rm log}~h$ in the
central $\sim 0.5~h^{-1}~{\rm Mpc}$ of the cluster.  CL0023+0423 has a
galaxy population which is more typical of groups and the field
population. The numbers from the statistical distribution are
consistent with almost all of the galaxies being normal spirals. The
spectral results support these numbers, independently indicating
spiral fractions of 66\% or more.  CL1604+4304, in contrast, has a
morphological composition which is characteristic of a normal,
present-day rich cluster. Early-type galaxies comprise 76\% of all
galaxies in this region. In this population, the ratio of S0 galaxies
to ellipticals is $1.7 \pm 0.9$, consistent with galaxy populations
found in local clusters (Dressler 1980a).

\item The morphological results support the conclusions of the
dynamical analysis presented in Paper II. CL0023+0423 is apparently
two independent systems separated in velocity by $\sim 2900~{\rm
km~s^{-1}}$.  The velocity dispersions and implied masses indicate
that these systems are similar to local galaxy groups or poor
clusters. Though this may be a chance projection, the dynamical and
morphological evidence may indicate that we are seeing the merger of
two spiral-dominated galaxy groups (see Lubin, Postman \& Oke
1998a). The velocity histogram of CL1604+4304, on the other hand, is
consistent with a Gaussian distribution, implying that this system
formed at an earlier epoch and is already relaxed. The velocity
dispersion and implied mass of this system are consistent with an
Abell richness class 2 or 3 cluster.

\end{list}

\vskip 0.5cm

We thank the anonymous referee for his thorough review of this paper.
Alan Dressler, Chris Fassnacht, and Ian Smail are thanked for useful
discussions, comments, and material aids to this paper.  It is also a
great pleasure to thank Allan Sandage for his generous gift of time
and invaluable expertise to this project.  The W.M. Keck Observatory
is operated as a scientific partnership between the California
Institute of Technology, the University of California, and the
National Aeronautics and Space Administration.  It was made possible
by generous financial support of the W. M. Keck Foundation. LML
graciously acknowledges support from a Carnegie Fellowship.  This
research was supported in part by {\it HST} GO analysis funds provided
through STScI grant GO-06000.01-94A and {\it HST} Archival grant 7536.

% Include Tables 1 - 4

\newpage

%\renewcommand{\baselinestretch}{1}

% Table 1c. Notes to Table 1a-b

\begin{table}
\begin{center}
Table 1c : Notes on Parameters listed in Tables 1a--b
\end{center}
\begin{center}
\scriptsize
\begin{tabular}[h]{llll}
\hline
\hline
Column	& Parameter	   & Units	& Comments \nl
\hline	 					
1	& MDS ID	   &	      	& MDS object identifier\tablenotemark{a} \nl
2	& Model \#	   &		& MLE model number\tablenotemark{b} \nl
3	& $N_{tot}$	   &	        & Total number of pixels in selected region\tablenotemark{c} \nl
4	& $N_{pix}$	   &	        & Total number of pixels in selected region above 1-$\sigma$\tablenotemark{c} \nl
5	& $N_{fit}$	   &		& Number of parameters in MLE model\tablenotemark{d} \nl
6	& Chip \#	   & 		& WFPC2 CCD number\tablenotemark{e} \nl
7	& X		   & Pixels	& X coordinate of centroid of MLE Model\tablenotemark{f} \nl
8	& Y		   & Pixels	& Y coordinate of centroid of MLE Model\tablenotemark{f} \nl
9 	& Sky		   & Mag	& MLE model sky magnitude\tablenotemark{g} \nl
10	& $m_{tot}$	   & Mag	& Total MLE model magnitude\tablenotemark{h} \nl
11	& $\delta m_{tot}$ & Mag	& Error on above \nl
12	& $R_{\frac{1}{2}}$	   & Log(arcsec)	& Log half-light radius of MLE model\tablenotemark{i} \nl
13	& $\delta R_{\frac{1}{2}}$ & Log(arcsec)	& Error on above \nl
14	& PA		   & Radians	& Orientation of MLE model\tablenotemark{j} \nl
15	& $\delta {\rm PA}$& Radians	& Error on above \nl
16	& ${(\frac{b}{a})}_{D}$ &	& Disk axis ratio of MLE model\tablenotemark{k} \nl
17	& $\delta {(\frac{b}{a})}_{D}$ 	&	& Error on above \nl
28	& ${(\frac{b}{a})}_{B}$ &	& Bulge axis ratio of MLE model\tablenotemark{k} \nl
19	& $\delta {(\frac{b}{a})}_{B}$ 	&	& Error on above \nl
20	& $\frac{B}{B+D}$		&	& Bulge/(Disk+Bulge) luminosity ratio\tablenotemark{l} \nl
21	& $\delta \frac{B}{B+D}$	&	& Error on above \nl
22	& $({\frac{B}{D}})_{R_{\frac{1}{2}}}$&& Log ratio of Bulge/Disk half-light radius of MLE model \nl
23	& $\delta ({\frac{B}{D}})_{R_{\frac{1}{2}}}$&& Error on above \nl
24	& $SNRIL$	   &		& Log integrated signal-to-noise ratio \nl
25	& Class		   &		& Name of MLE classification\tablenotemark{m} \nl
26	& A		   &		& Asymmetry parameter\tablenotemark{n} \nl
\hline

\tablecomments{Machine-readable forms of Tables 1a--b can be obtained 
from the website address http://landru.stsci.edu:5000/hizclus/ftp.html
(see Sect.\ 3.1.1).  In addition, all of the data from the MDS
reduction (including the processing information, the raw and reduced
images, and the full catalogs) are publically available from the STScI
archive. The data for both cluster fields can be found at
http://archive.stsci.edu/mds/mds.cgi.  Here, one must first choose the
option ``Define Fields'' to specify the fields of interest by an
RA/Dec range and the minimum number of exposures in a given
passband. The option ``Find Fields'' will then retrieve the specified
observations. The CL0023+0423 observations are designated dataset
u2vk1, and the CL1604+4304 observations are designated datasets u2845
and u2845.} \nl

\tablenotetext{}{}

\tablenotetext{a}{Original identification number from the MDS reduction
pipeline. Those numbers which are excluded indicate detected objects
which were classified as stars or which did not reach the required
signal-to-noise (see Sect.\ 3.1.1).}

\tablenotetext{b}{MDS model number : 1, disk; 2, bulge; 3, disk+bulge.}

\tablenotetext{c}{$N_{tot}$ is the total number of usable pixels in the
selected 64 or 128--pixel square region around each object; $N_{pix}$
is the number of $N_{tot}$ pixels which are more than 1-$\sigma$ above
the estimated local sky background (see Sect.\ 3.1.1).}

\tablenotetext{d}{The number of parameters in the fit which actually vary. A
maximum of 12 parameters are fit, though in most cases one or more
parameters are held fixed (see e.g.\ table note m).}

\tablenotetext{e}{CCD chip number : PC = 1; WFC = 2,3,4.}

\tablenotetext{f}{The mean error between the model centroid and the actual
centroid of the object image is 0.2 pixels. Coordinates are relative
to each individual chip.}

\end{tabular}
\end{center}
\end{table}

\newpage

\begin{table}
\begin{tabular}[h]{llll}
\end{tabular}

\tablenotetext{g}{A maximum likelihood estimate for the local sky background
is determined simultaneously with the other model
parameters. Sufficient pixels are used to ensure that the sky level is
determined to an accuracy of 0.5\%.  The sky background is assumed to
be flat over the small region selected for analysis.}

\tablenotetext{h}{Small differences between the analytic total magnitude
and the true total magnitude may arise because the true galaxy is not
smooth, and the model may not average properly over bright regions of
star formation; for details, see references listed in text.}

\tablenotetext{i}{The radius within which half of the light of the
unconvolved model would be contained if it were radially symmetric (an
axis ratio of unity). Lower and upper limits of 0.1 pixel and one-half
of the maximum radius of the region selected for analysis have been
imposed. The half-light radius of the individual components can be
derived used the Bulge/Disk half-light radius ratio (column 22).}

\tablenotetext{j}{The adopted position angle (measured East from North) is
that of the axis of symmetry of the model. It is set equal to zero
when the source is assumed to be azimuthally symmetric with an axis
ratio of unity. For disk+bulge models, the orientation is assumed to
be the same for both components.}

\tablenotetext{k}{The ratio of the minor axis half-light ratio to that of the
major axis. For disk+bulge models it is defined independently for each
component. If the axis ratio is not significantly different from
unity, it is fixed at 1. A ratio of 0 indicates that this component
was not fit to the data.}

\tablenotetext{l}{The luminosity of the individual components can be
derived from this parameter and the total magnitude $m_{tot}$ (column
10).}

\tablenotetext{m}{Object classification : object, galaxy, disk, bulge or
disk+bulge.}

\tablenotetext{n}{A value of 99.99 denotes an undefined measurement.}

\end{table}

% Table 2a : CL0023+0423

\newpage

\textheight=9.0in
\textwidth=7.0in
\voffset -0.75in
\hoffset -0.5in



% Table 2c : Notes to Table 2a-b

\clearpage

\begin{deluxetable}{cll}
\tablewidth{0pt}
\tablenum{2c}
\scriptsize
\tablecaption{Notes on Parameters listed in Tables 2a--b}
\tablehead{
\colhead{Column}   &
\colhead{Heading}  &
\colhead{Comments}}
\startdata
1	&	ID		&	MDS object identifier\tablenotemark{a} \nl
2	&	$m_{tot}$	&	Analytic total magnitude\tablenotemark{b} \nl
3	&	Class		&	Hubble class\tablenotemark{c} \nl
4	&	D		& 	Disturbance index\tablenotemark{d} \nl
5	&	Interp		&	Interpretation of the disturbance\tablenotemark{e} \nl
6	&	Comments	&	Description of object morphology\tablenotemark{f} \nl

\tablenotetext{a}{MDS indentification number as given in Table 1.}

\tablenotetext{b}{Total magnitude of best-fit analytic model in either
$R_{702}$ (CL0023+0423) or $I_{814}$ (CL1604+4304) as given in Table
1.}

\tablenotetext{c}{The standard Hubble classification scheme (E, S0, Sa, Sab, Sb...)
with the addition of the following : E/S0 or S0/E, cannot distinguish
between E or S0; X, compact object (likely non-stellar, but too
compact to see structure); P, peculiar or unclassifiable.}

\tablenotetext{d}{Disturbance index : 0, normal; 1, moderate asymmetry; 2,
strong asymmetry; 3, moderate distortion; 4, strong distortion.}

\tablenotetext{e}{Interpretation of disturbance index : M, merger; I, tidal
interaction with neighbor; T, tidal feature; C, chaotic.}

\tablenotetext{f}{Here, comp is short for ``companion''; @ indicates position 
relative to the galaxy going clockwise, e.g.\ @7 = ``at 7 o'clock.''}

\enddata
\end{deluxetable}

% Table 3 of Comparison between Visual and Automated Typing

\begin{table}
\begin{center}
Table 3. Comparison Between Visual and Automated Typing
\end{center}
\begin{center}
%

% Include Figure Files 1 & 2

\def\plottwo#1#2{\centering \leavevmode
    \epsfxsize=.5\columnwidth \epsfbox{#1} \hfil
    \epsfxsize=.5\columnwidth \epsfbox{#2}}

% set up page limit specification

\textheight=8.5in
\textwidth=6.5in
\voffset 0.0in
\hoffset 0.0in

\clearpage

\begin{figure}
%\centerline{\epsfbox{00final2.ps}}
PLATE \#1
\caption{The WFPC2 image of cluster field CL0023+0423. The total
exposure time is 17.9 ksec through the F702W filter. The X's indicate
the centers of the two substructures in this system as determined from
the dynamical analysis of the Keck spectroscopic data over the full
LRIS field-of-view (see Sect.\ 3 of Paper II).}
\label{00hst}
\end{figure}

\begin{figure}
%\centerline{\epsfbox{16mdsfinal95.ps}}
PLATE \#2
\caption{The WFPC2 image of cluster field CL1604+4304. The total
exposure time is 32.0 ksec through the F814W filter. The X indicates
the dynamical center of the cluster as determined from the dynamical
analysis of the Keck spectroscopic data over the full LRIS
field-of-view (see Sect.\ 3 of Paper II).}
\label{16hst}
\end{figure}

\begin{figure}
\centerline{\epsfbox{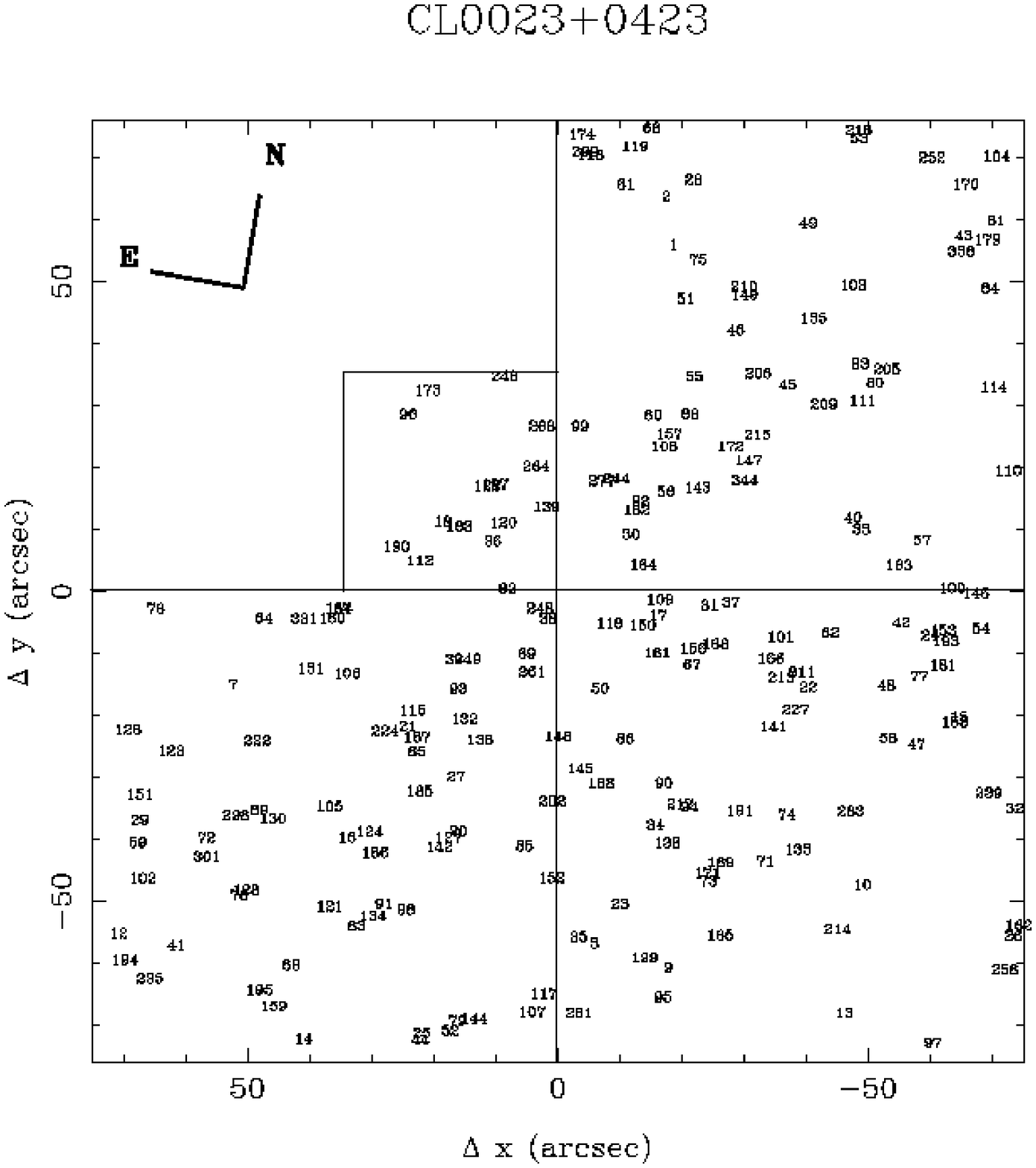}}
\caption{The finding chart for the visually-classified subsample of
galaxies in the WFPC2 image of cluster field CL0023+0423 (see Sect.\
3.2).  The MDS identification number is indicated such that the lower
left corner of the first digit is at the position of each galaxy. This
figure can be used as an overlay with Figure~\ref{00hst}.}
\label{00index}
\end{figure}

\begin{figure}
\centerline{\epsfbox{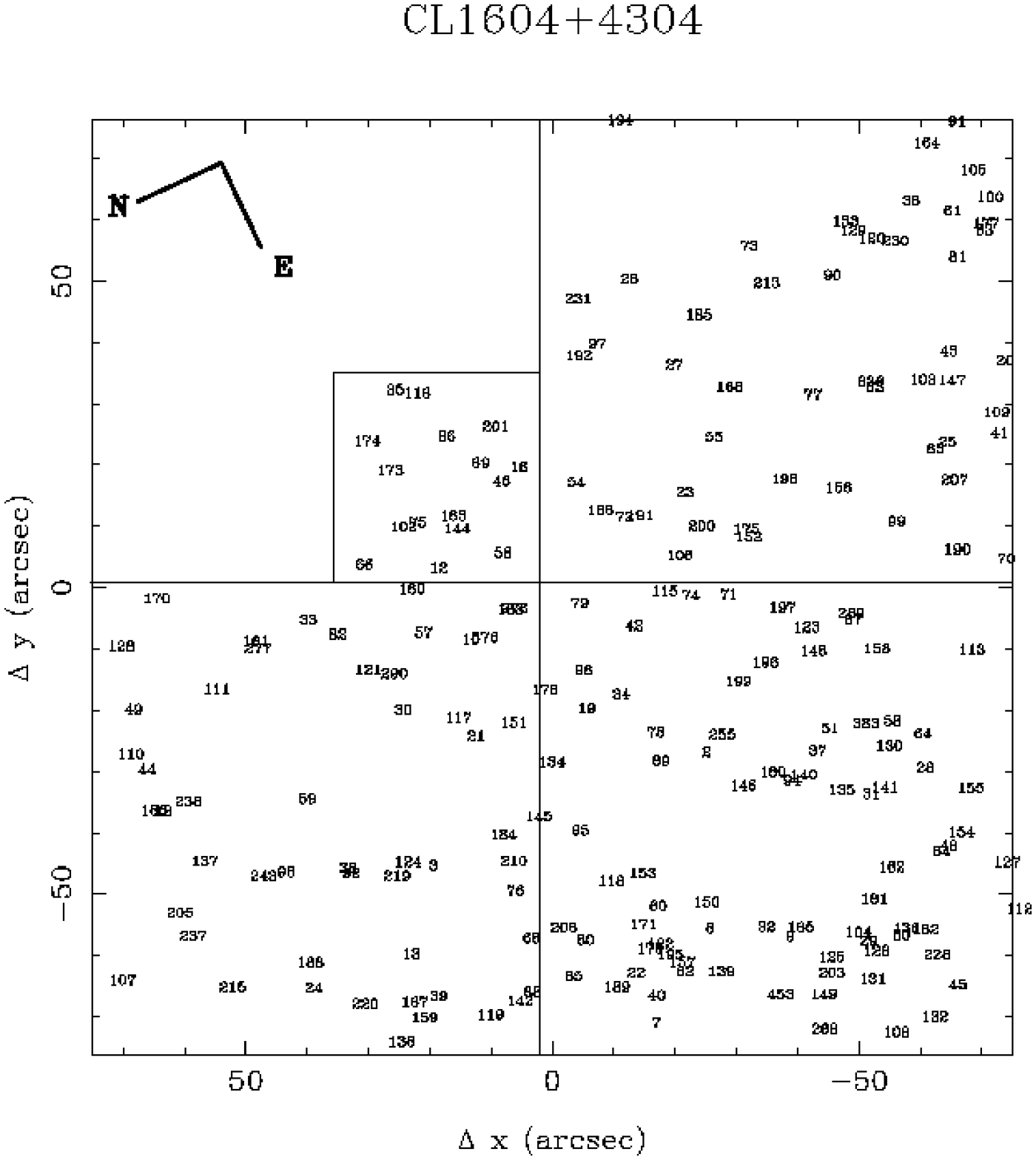}}
\caption{The finding chart for the visually-classified subsample of
galaxies in the WFPC2 image of cluster field CL1604+4304 (see Sect.\
3.2).  The MDS identification number is indicated such that the lower
left corner of the first digit is at the position of each galaxy.
This figure can be used as an overlay with Figure~\ref{16hst}.}
\label{16index}
\end{figure}

\begin{figure}
\epsfysize=5.0in
\centerline{\epsfbox{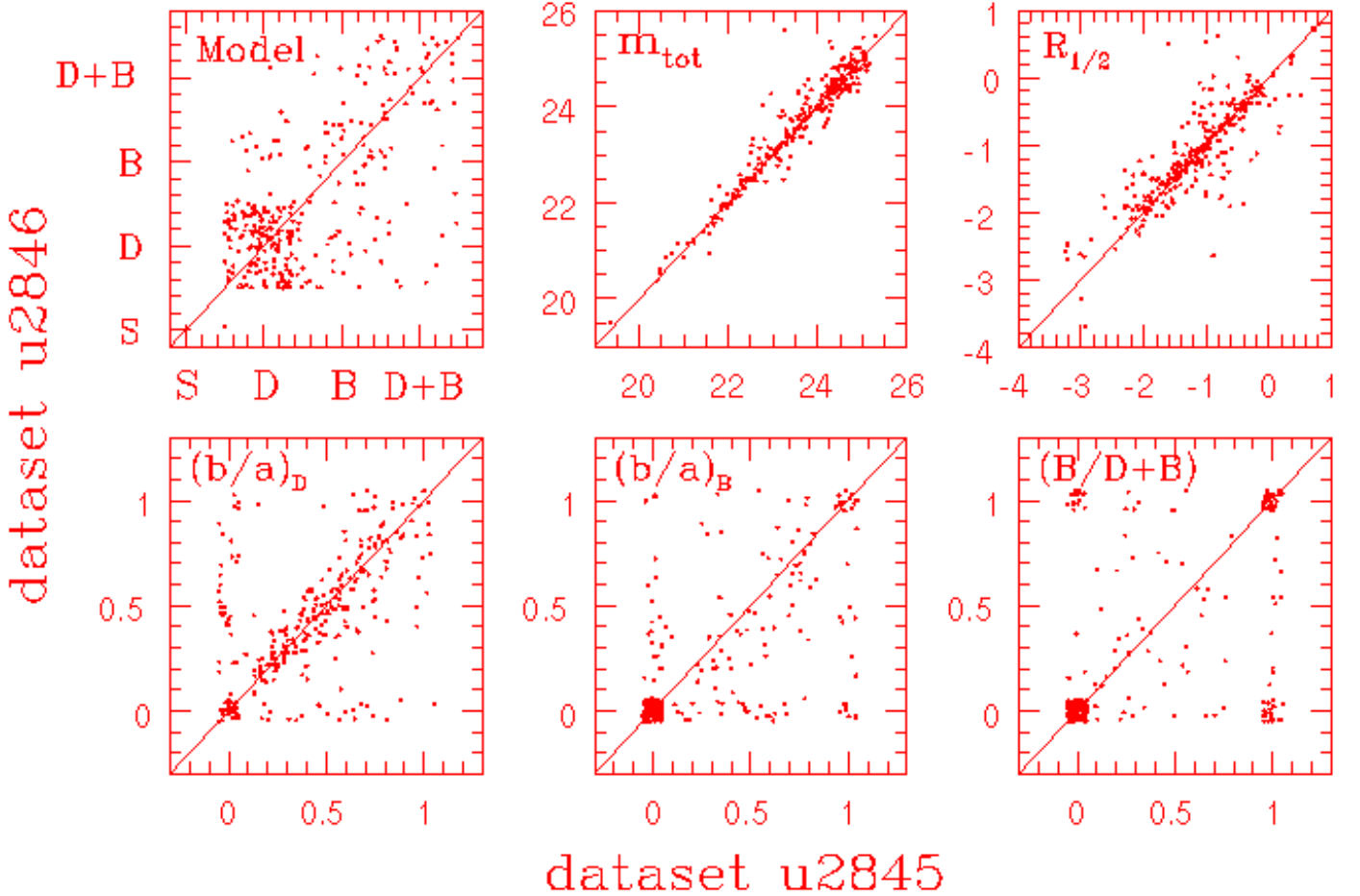}}
\caption{A comparison of the model parameters for the same galaxy derived
from the two separate datasets, u2845 and u2846, of the CL1604+4304
observations. Only those galaxies with $SNRIL > 2$ have been used in
this analysis (see Sect.\ 3.1.1). {\it Upper} : The left panel
indicates the best-fit model, either star (S), disk (D), bulge (B), or
disk+bulge (D+B). Each data point has been offset randomly by +/- half
a class unit to show the true density of points. The middle panel
shows the analytic total magnitude $m_{tot}$ in Vega magnitudes. The
right panel shows the logarithm of the half-light radius
$R_{\frac{1}{2}}$ in Log(arcsec). {\it Lower} : The left and middle
panels show the axial ratio of the disk ${(\frac{b}{a})}_{D}$ and
bulge ${(\frac{b}{a})}_{B}$ components, respectively. The right panel
indicates the bulge/(disk+bulge) luminosity ratio $\frac{B}{D+B}$. The
line in each panel indicates an equality between the u2845 and u2846
values. The median of the ratio $\frac{u2845}{u2846}$ for each of
these parameters is consistent with 1 (see Sect.\ 3.1.1).}
\label{mdscomp}
\end{figure}

\begin{figure}
\epsfysize=6.5in
\centerline{\epsfbox{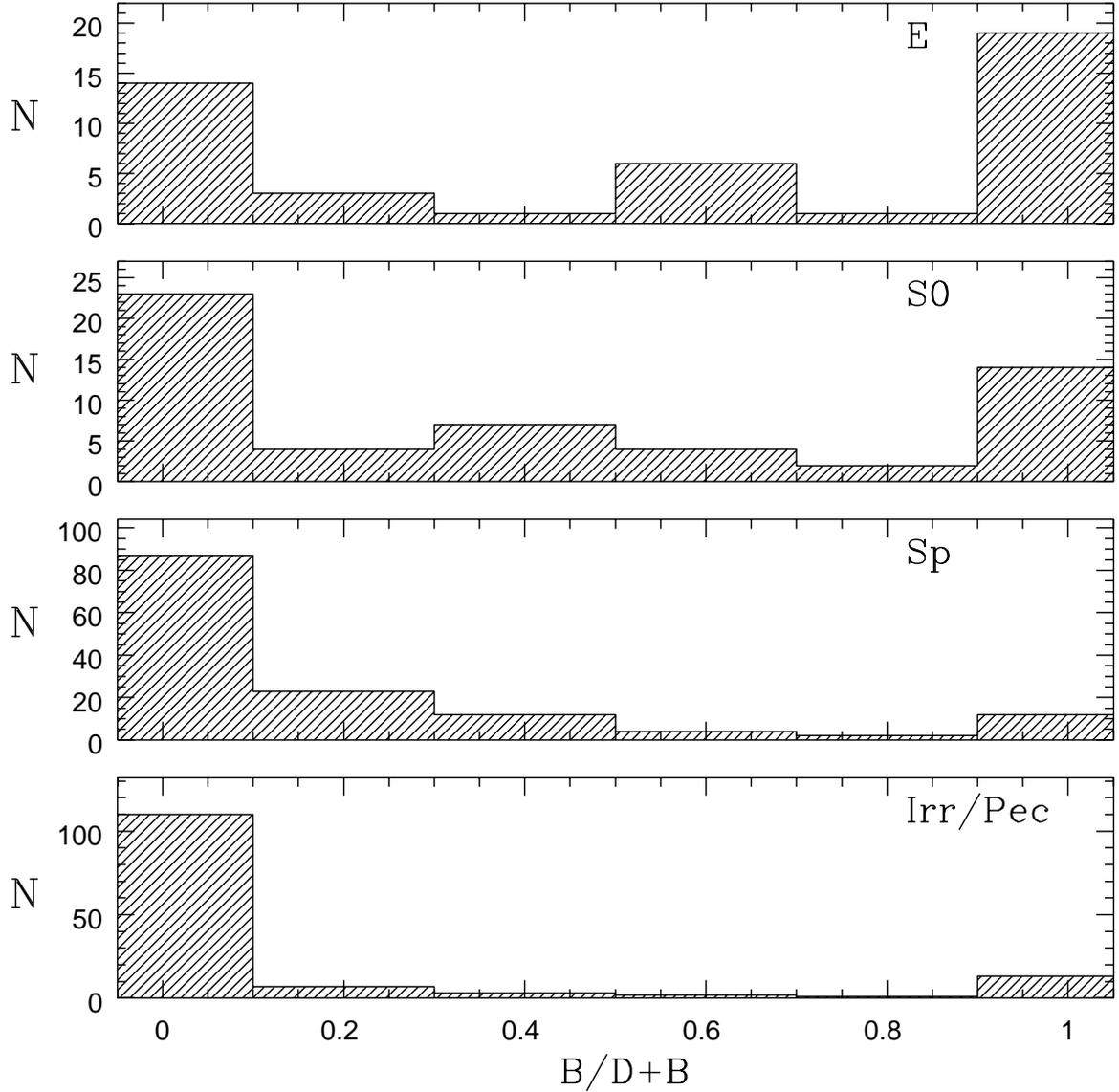}}
\caption{Distribution of the bulge/(disk+bulge) luminosity ratios
for different morphological types from the combined data of both
cluster fields. Early-type galaxies have a distribution of
$\frac{B}{D+B}$ luminosity ratios which are more bi-modal between 0
(pure disk) and 1 (pure bulge), whereas spiral and irregular/peculiar
galaxies clearly are weighted much more heavily to ratios of 0 (see
Sect.\ 4).}
\label{mdsbtt}
\end{figure}

\begin{figure}
\epsfysize=6.5in
\centerline{\epsfbox{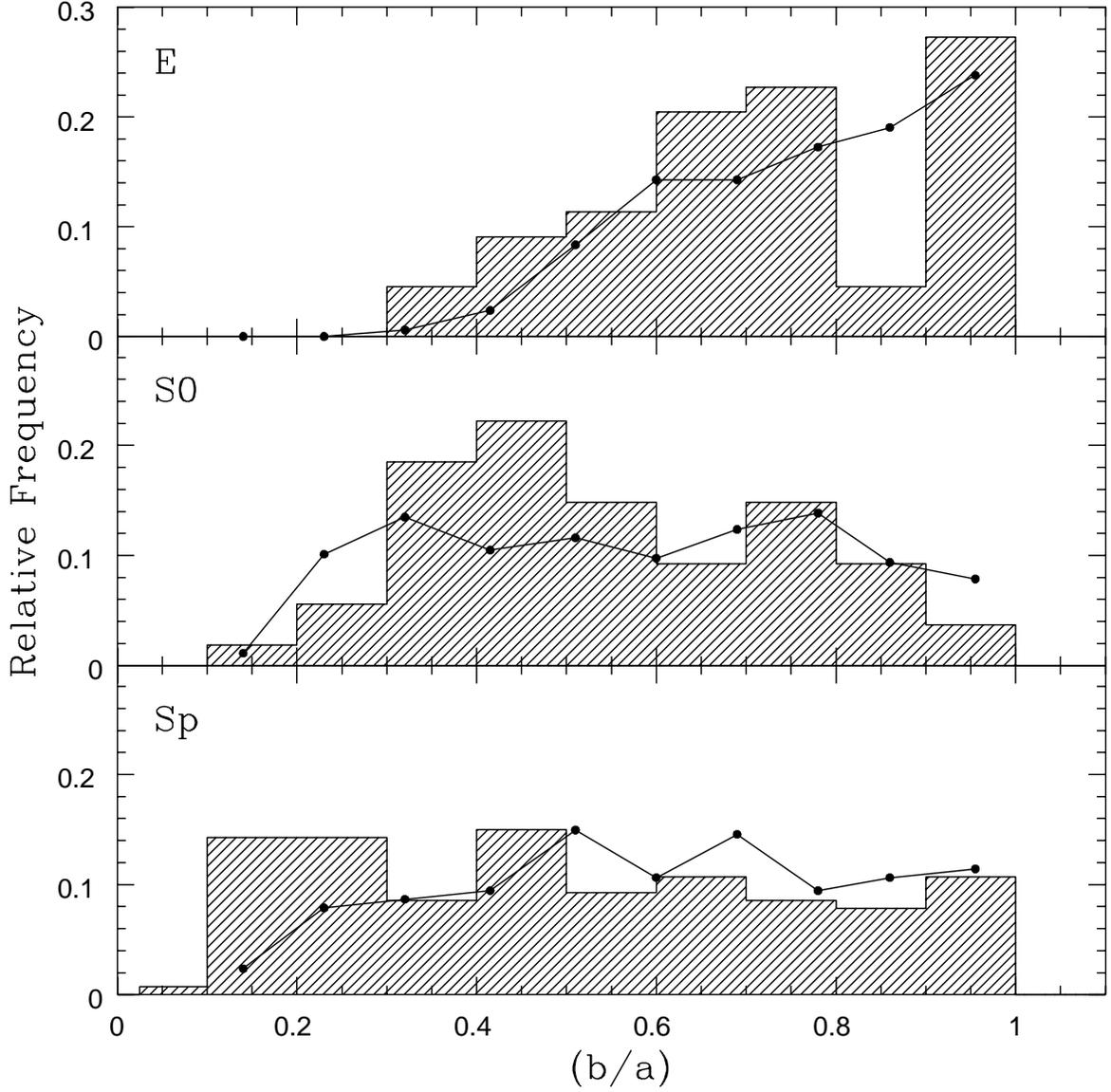}}
\caption{Distribution of the axial ratios ${(\frac{b}{a})}$ from different
morphological types for the combined data of both cluster fields. The
axial ratio of the best-fit model is used.  In the cases where the
best-fit model is a disk+bulge model, the axial ratio of the brightest
component (disk or bulge) is assumed. The curves indicate the
distribution of axial ratios from a sample of nearby galaxies
(Sandage, Freeman \& Stokes 1970). In each case, the distribution of
our distant sample is consistent with the distribution of the nearby
sample (see Sect.\ 4).}
\label{mdsab}
\end{figure}

\begin{figure}
\epsfysize=6.5in
\centerline{\epsfbox{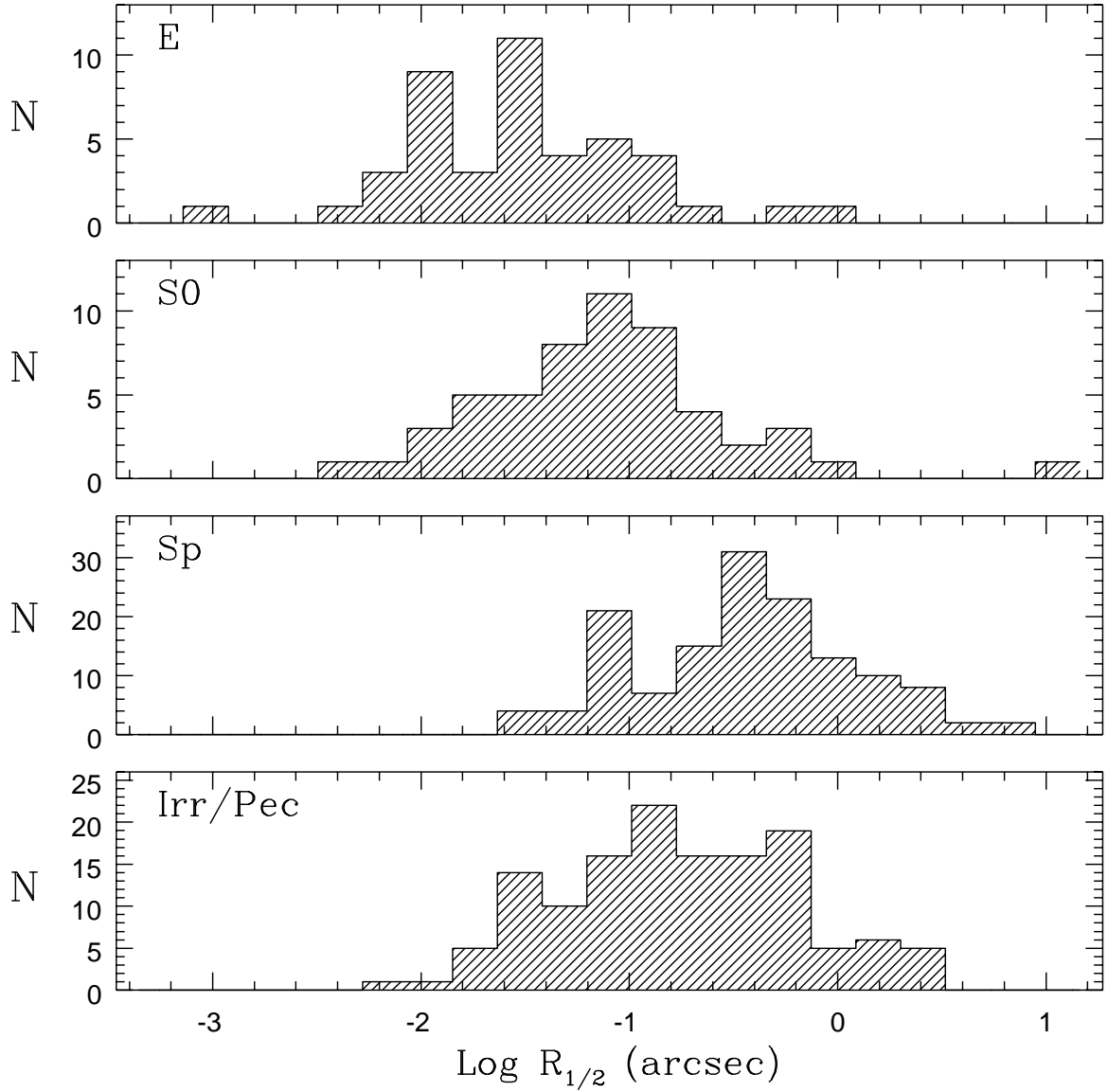}}
\caption{Distribution of the logarithm of the half-light radius
($R_{\frac{1}{2}}$) for different morphological types from the
combined data of both cluster fields. There is clearly a progression
with morphological type. Early-type galaxies have smaller half-light
radii than late-type galaxies.}
\label{mdshlrg}
\end{figure}

\begin{figure}
\epsfysize=6.5in
\centerline{\epsfbox{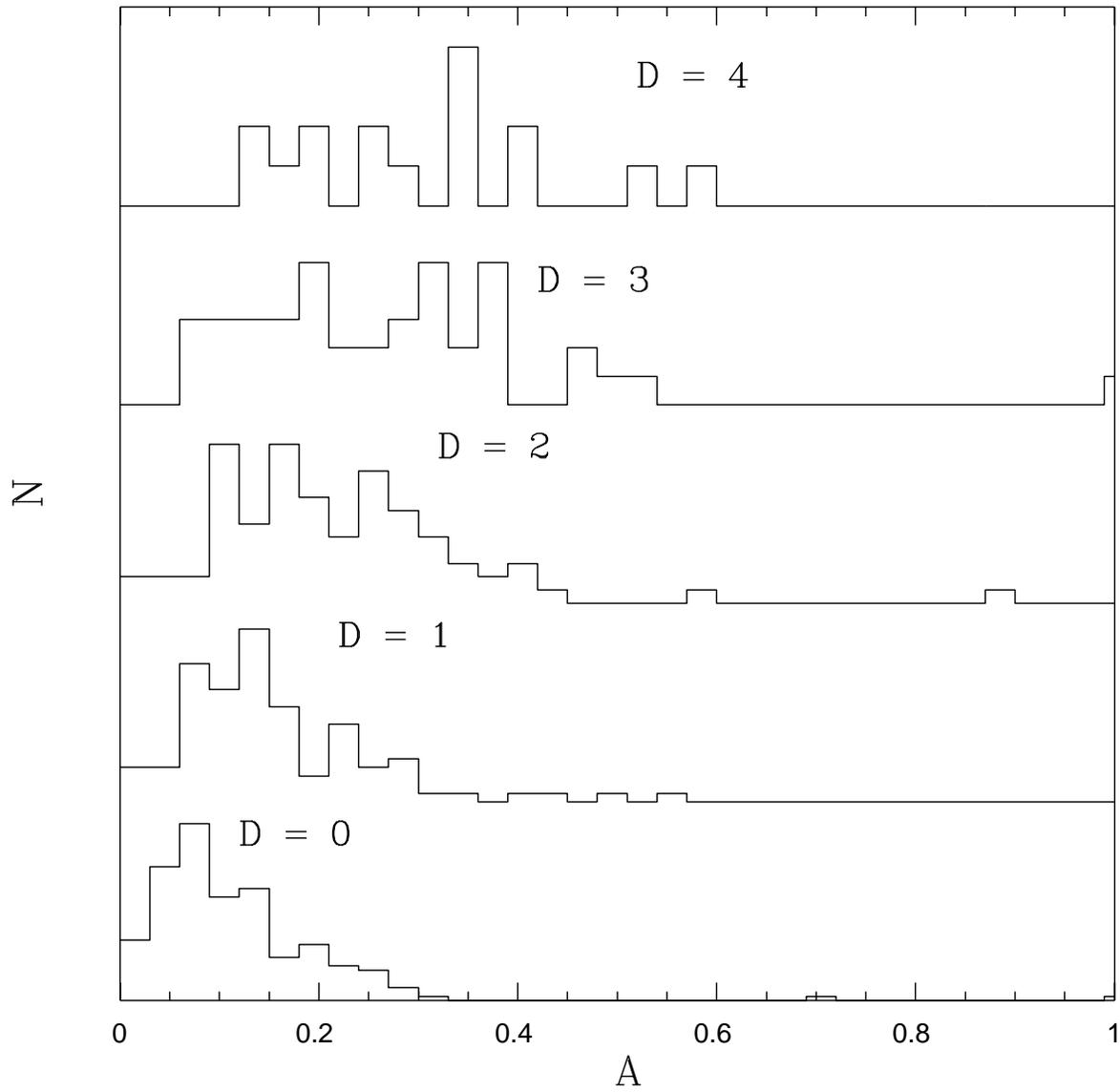}}
\caption{Distribution of asymmetry parameter ($A$) for different disturbance
indices ($D$) from the combined data of both cluster fields. There
appears to be a correlation between the automated and visual
parameters. }
\label{arescomp}
\end{figure}

\begin{figure}
\plotone{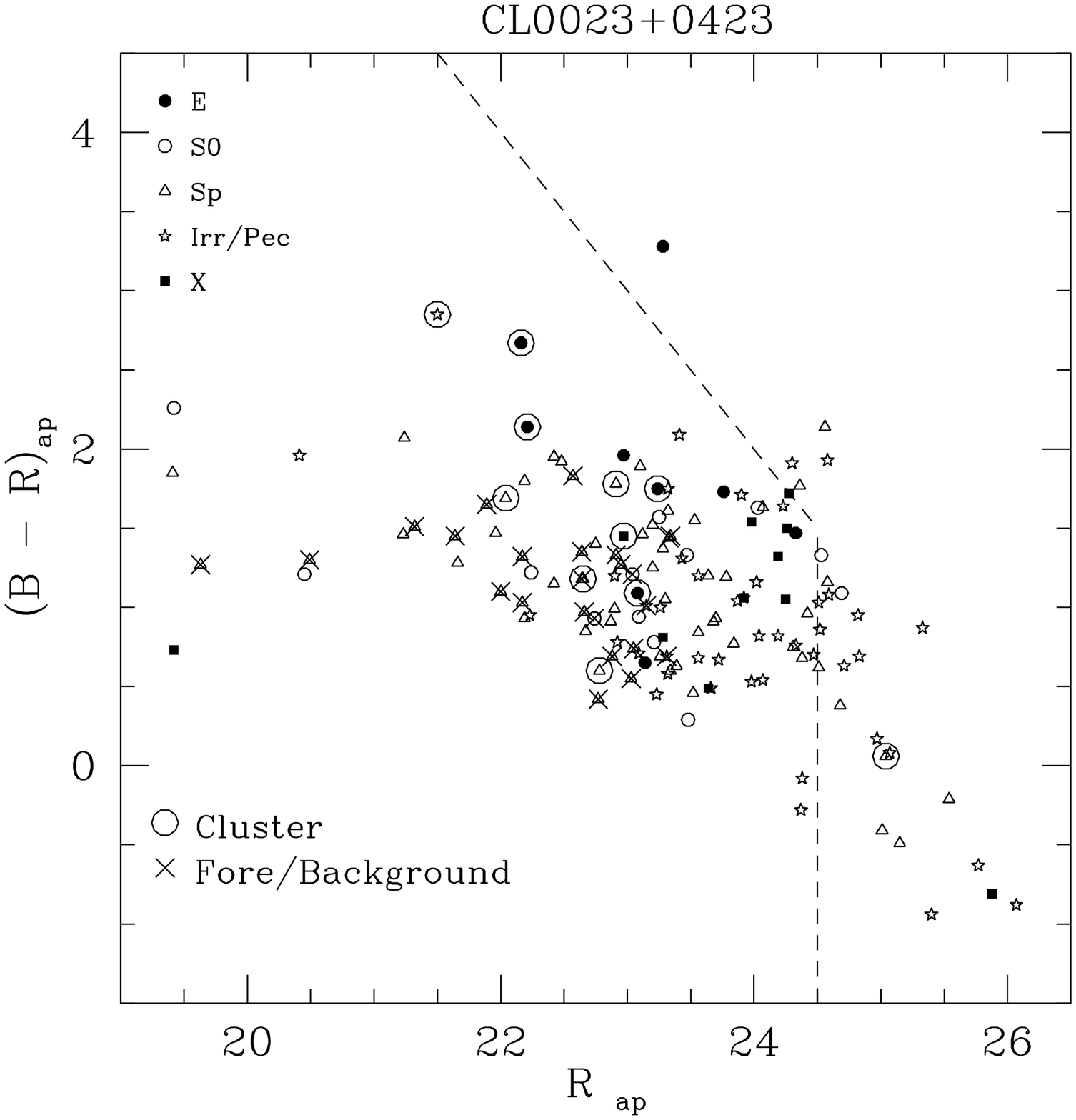}
\caption{The $B-R$ versus $R$ color-magnitude diagrams of the cluster
field CL0023+0423. The magnitudes are calculated within an aperture of
radius $3^{''}$.  The galaxy morphology is indicated by different
symbols : filled circle -- elliptical, open circle -- S0, triangle --
spiral, star -- irregular, and filled square -- compact. The galaxies
with measured redshifts are circled if they are a cluster member (as
determined from the velocity analysis in Sect.\ 3 of Paper II) or
crossed out if they are foreground or background. The magnitude limits
of the Keck photometric survey are indicated by the dashed lines (see
Sect.\ 2.2.1).}
\label{00brmorph}
\end{figure}

\begin{figure}
\plotone{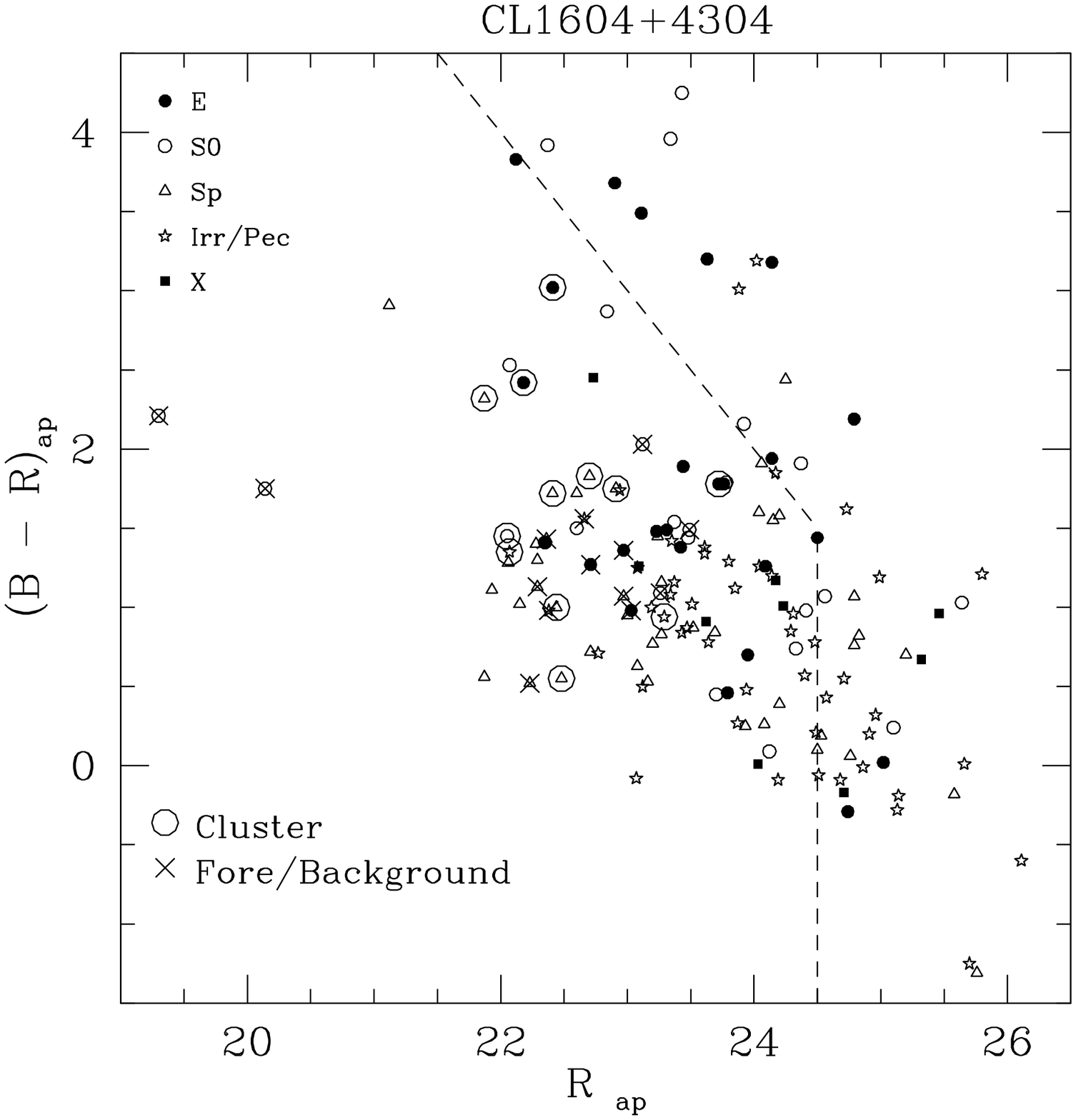}
\caption{The $B-R$ versus $R$ color-magnitude diagrams of the cluster
field CL1604+4304. The magnitudes are calculated within an aperture
of radius $3^{''}$.  The symbols are the same as in
Figure~\ref{00brmorph}.  The magnitude limits of the Keck photometric
survey are indicated by the dashed lines (see Sect.\ 2.2.1).}
\label{16brmorph}
\end{figure}

\begin{figure}
\plotone{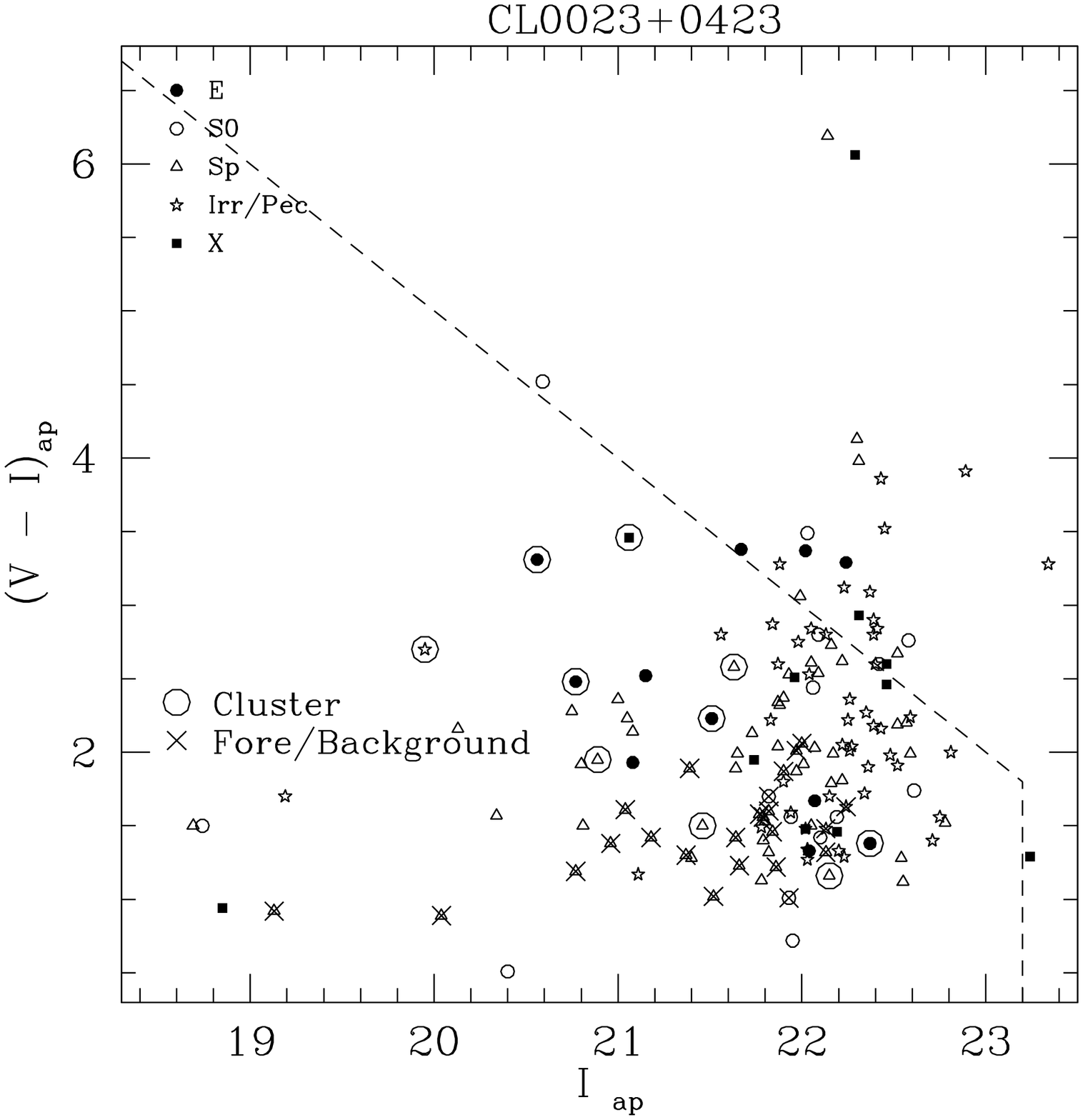}
\caption{The $V-I$ versus $I$ color-magnitude diagrams of the cluster
field CL0023+0423. The magnitudes are calculated within an aperture of
radius $3^{''}$.  The galaxy morphology is indicated by different
symbols : filled circle -- elliptical, open circle -- S0, triangle --
spiral, star -- irregular, and filled square -- compact. The galaxies
with measured redshifts are circled if they are a cluster member (as
determined from the velocity analysis in Sect.\ 3 of Paper II) or
crossed out if they are foreground or background. The magnitude limits
of the Keck photometric survey are indicated by the dashed lines (see
Sect.\ 2.2.1).}
\label{00vimorph}
\end{figure}

\begin{figure}
\plotone{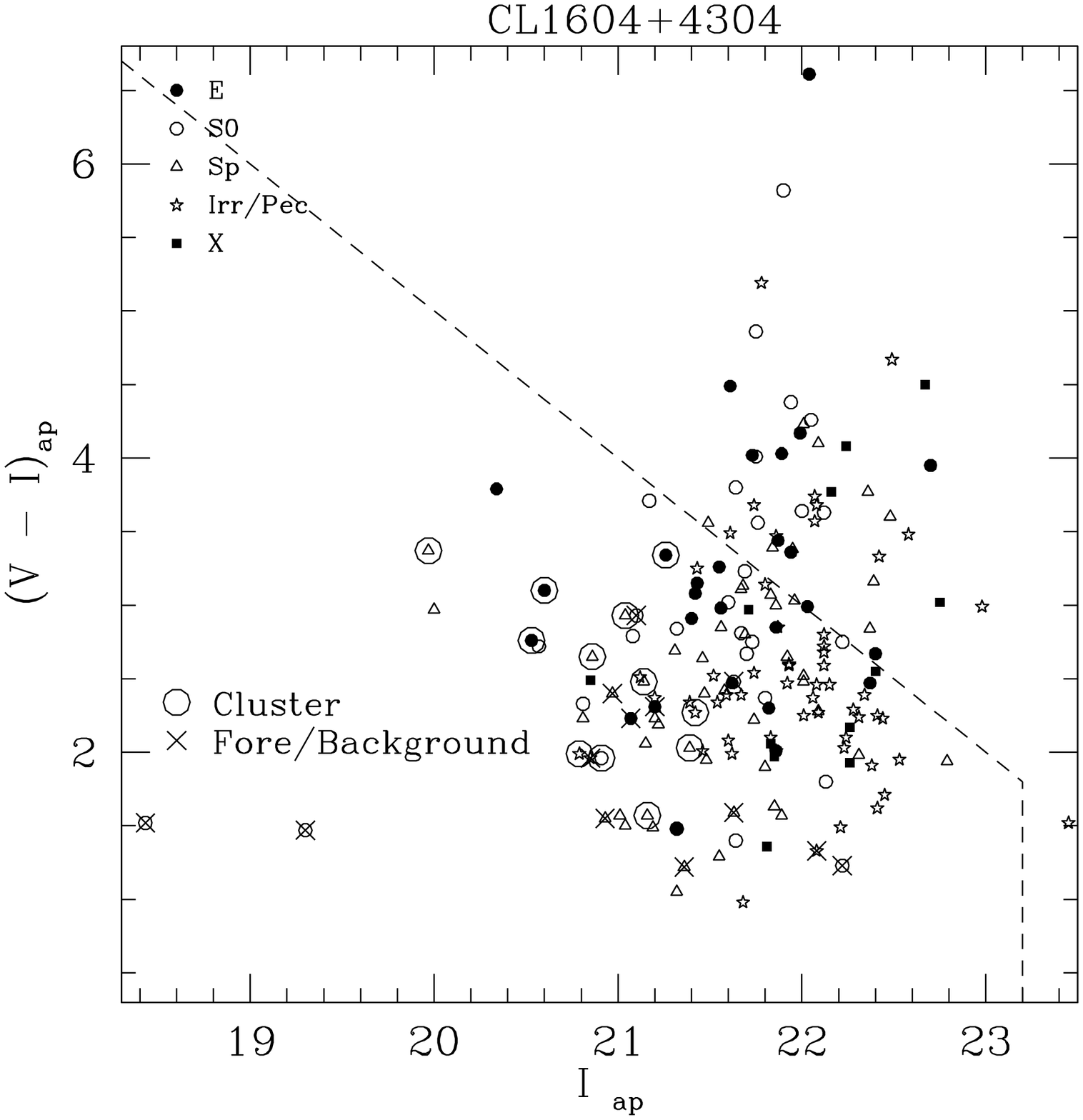}
\caption{The $V-I$ versus $I$ color-magnitude diagrams of the cluster
field CL1604+4304. The magnitudes are calculated within an aperture of
radius $3^{''}$. The symbols are the same as in
Figure~\ref{00vimorph}. The magnitude limits of the Keck photometric
survey are indicated by the dashed lines (see Sect.\ 2.2.1).}
\label{16vimorph}
\end{figure}

\begin{figure}
\epsfysize=6.0in
\centerline{\epsfbox{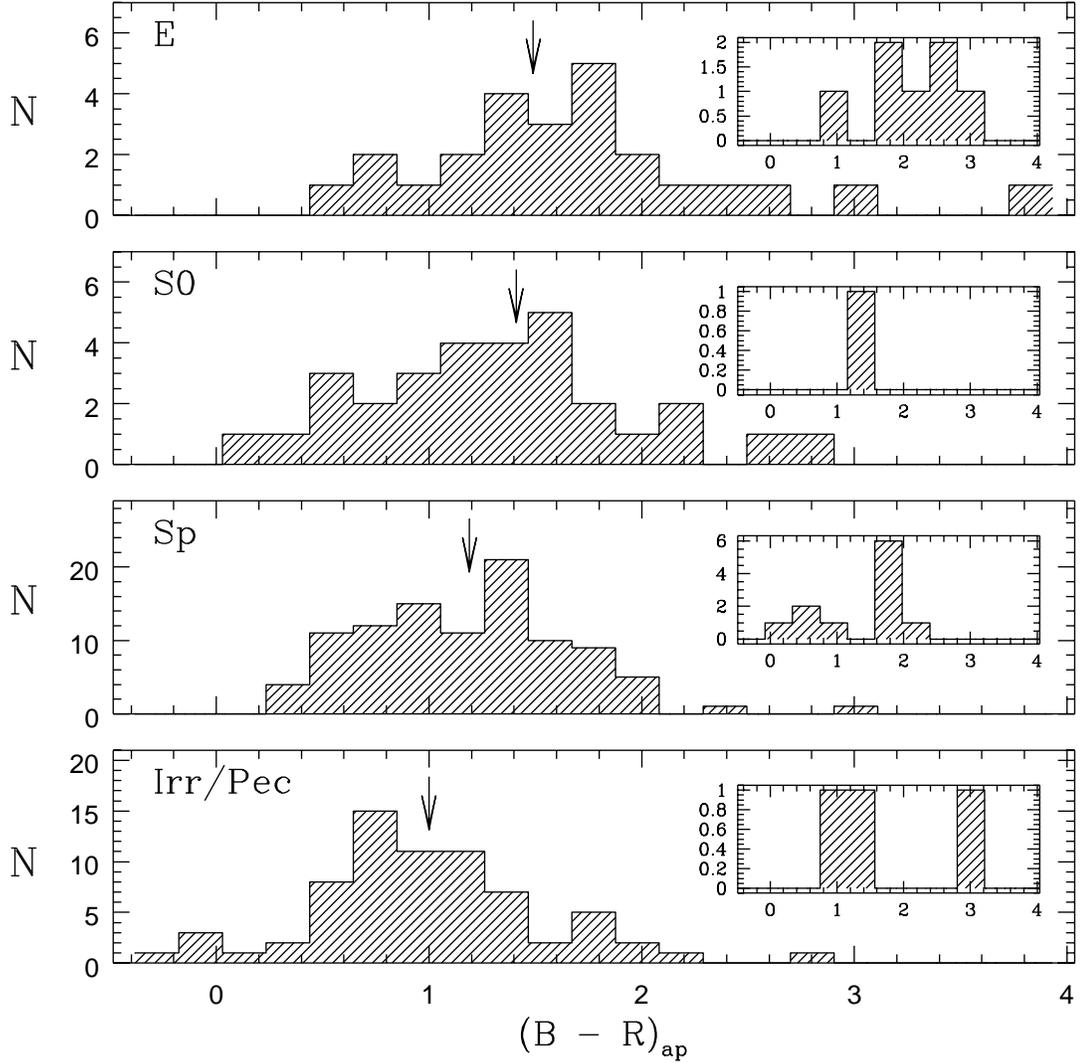}}
\caption{Distribution of $(B-R)$ colors for various morphological
types. Here, we use only those galaxies that are within the magnitude
limits of the sample. The arrow in each panel indicates the median
color of the distribution. There is a clear progression in color from
early- to late-type galaxies. As expected, elliptical and S0s are
redder, on average, than spirals and irregulars/peculiars (see Sect.\
5.1). The inset window in each panel shows the $(B-R)$ color
distribution of only the confirmed cluster members from the combined
CL0023+0423 and CL1604+4304 fields.}
\label{brdist}
\end{figure}

\begin{figure}
%\centerline{\epsfbox{00hst_z_im.ps}}
PLATE \#3
\caption{All of the galaxies in the HST image of cluster field CL0023+0423
with measured redshifts from the Keck observations. The cluster
galaxies are shown in the top section of the figure, while the field
galaxies are shown in the bottom section. In each section, the
galaxies are ordered according to increasing redshift. The
field-of-view of each panel is $5\farcs{98} \times 5\farcs{98}$. The
redshift is given in the upper left corner of each panel. The two
numbers at the bottom of each panel indicate the Keck photometric
identification number and the MDS object identification number,
respectively. In four cases, the object was not detected in the
automated MDS analysis and, therefore, has no MDS identification
number (see Sect.\ 5.2).}
\label{00hstz}
\end{figure}

\begin{figure}
%\centerline{\epsfbox{16hst_z_im.ps}}
PLATE \#4
\caption{All of the galaxies in the HST image of cluster field CL1604+4304
with measured redshifts from the Keck observations. The cluster
galaxies are shown in the top section of the figure, while the field
galaxies are shown in the bottom section. In each section, the
galaxies are ordered according to increasing redshift. The
field-of-view of each panel is $5\farcs{98} \times 5\farcs{98}$. The
redshift is given in the upper left corner of each panel. The two
numbers at the bottom of each panel indicate the Keck photometric
identification number and the MDS object identification number,
respectively. In two cases, the object was not detected in the
automated MDS analysis and, therefore, has no MDS identification
number (see Sect.\ 5.2).}
\label{16hstz}
\end{figure}

\begin{figure}
\epsfysize=6.5in
\centerline{\epsfbox{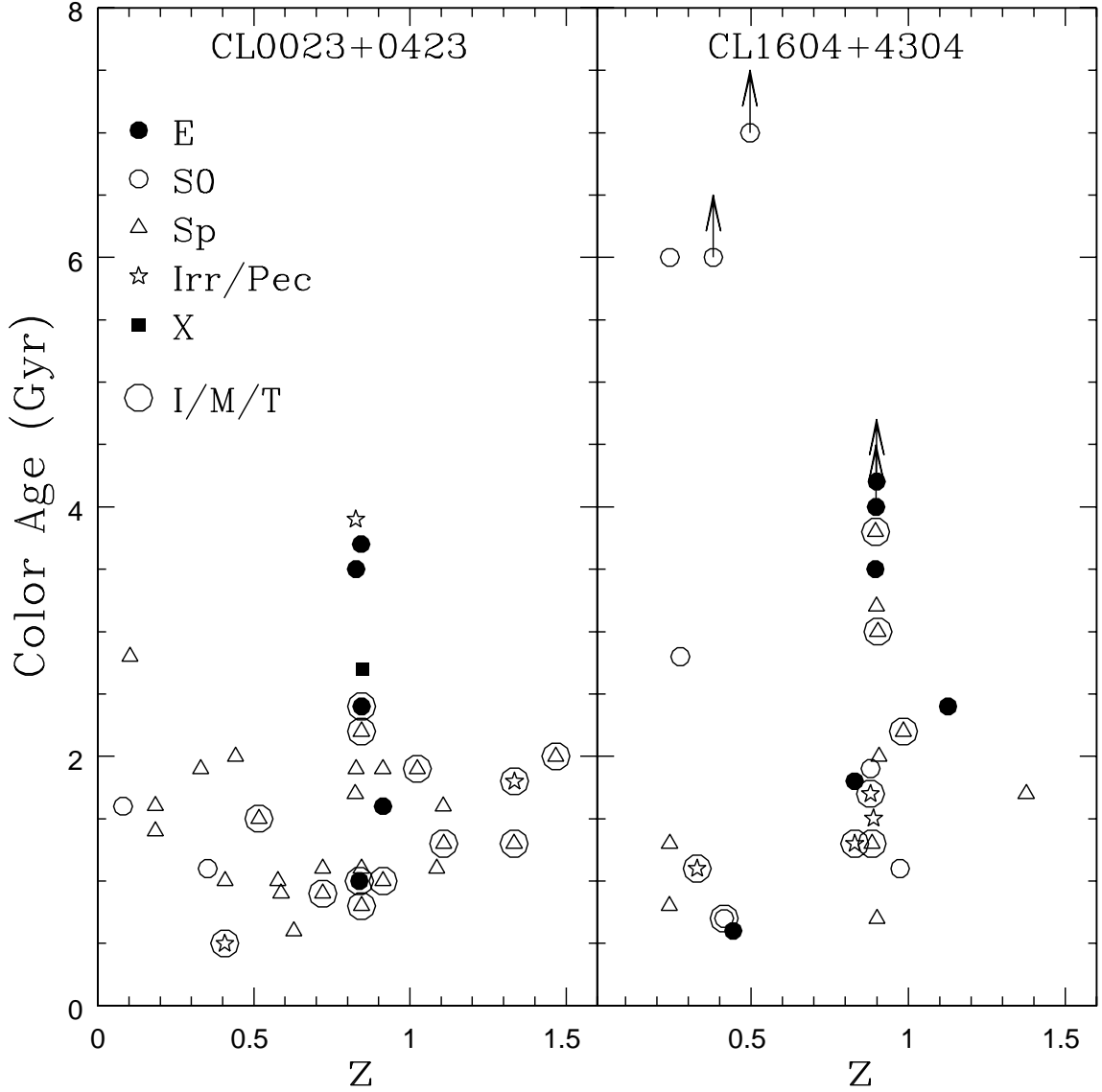}}
\caption{Color ages of HST galaxies versus redshift for the each cluster
field. The ages refer to the time since the last period of major star
formation and are determined through a comparison between the galaxy
spectral energy distribution and the Bruzual \& Charlot $\tau = 0.6$
Gyr model. Ages which are lower limits are indicated by arrows (see
Sect.\ 5.2). The galaxy morphology is indicated by different symbols :
filled circle -- elliptical, open circle -- S0, triangle -- spiral,
star -- irregular, and filled square -- compact. Galaxies with
evidence of an interaction (I), merger (M), or tidal feature (T) are
circled. The cluster redshift is obvious in both panels.}
\label{agez}
\end{figure}

\clearpage

\begin{figure}
\epsfysize=6.5in
\centerline{\epsfbox{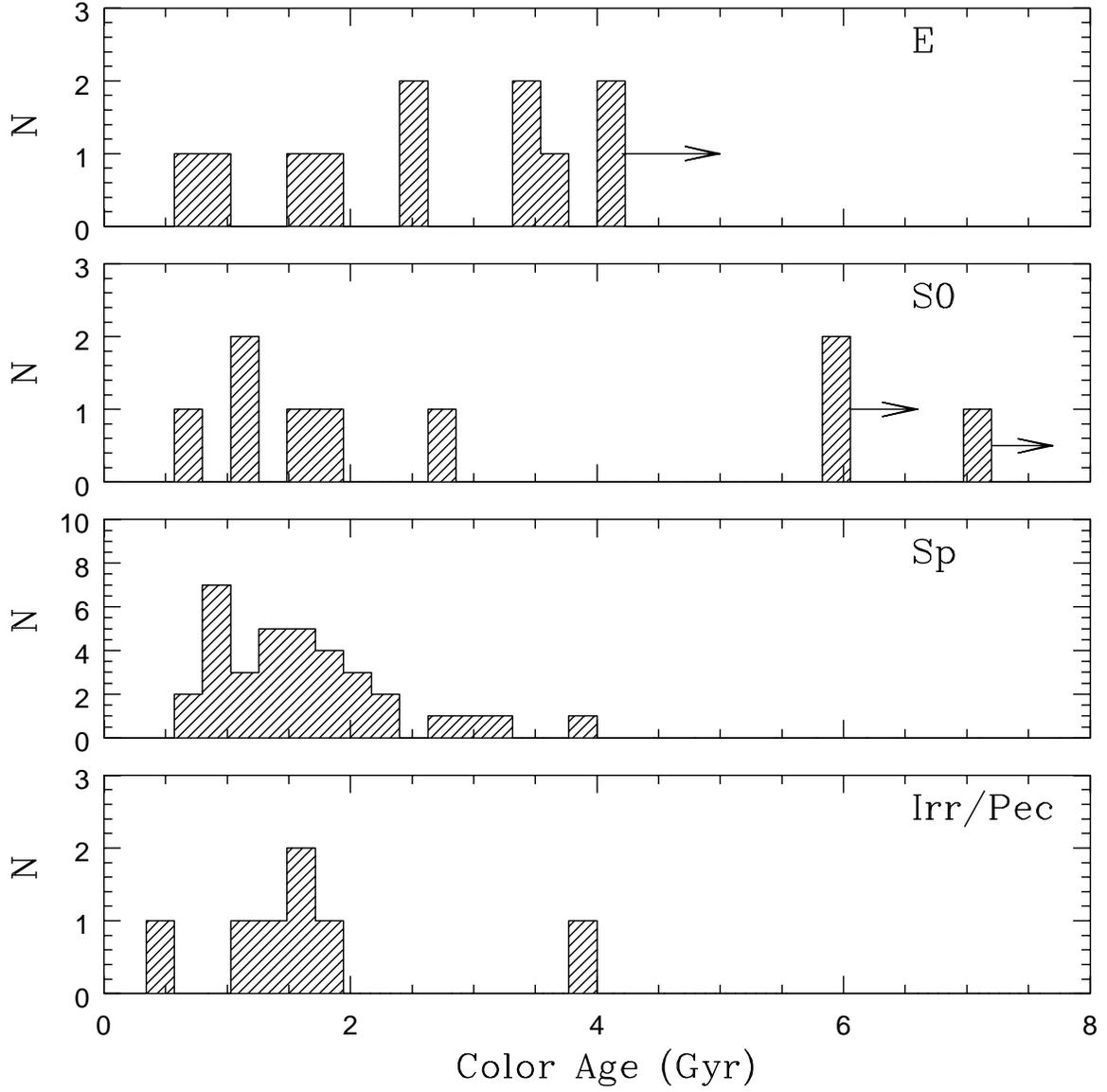}}
\caption{Distribution of color ages for different morphological types for
the combined CL0023+0423 and CL1604+4304 HST fields.  The ages refer
to the time since the last period of major star formation and are
determined through a comparison between the galaxy spectral energy
distribution and the Bruzual \& Charlot $\tau = 0.6$ Gyr model (see
Sect.\ 5.2). Ages which are lower limits are indicated by arrows. The
binning of 0.2 Gyr roughly corresponds to a typical error in the color
ages (see Paper II).}
\label{agehist}
\end{figure}

\begin{figure}
\centerline{\epsfbox{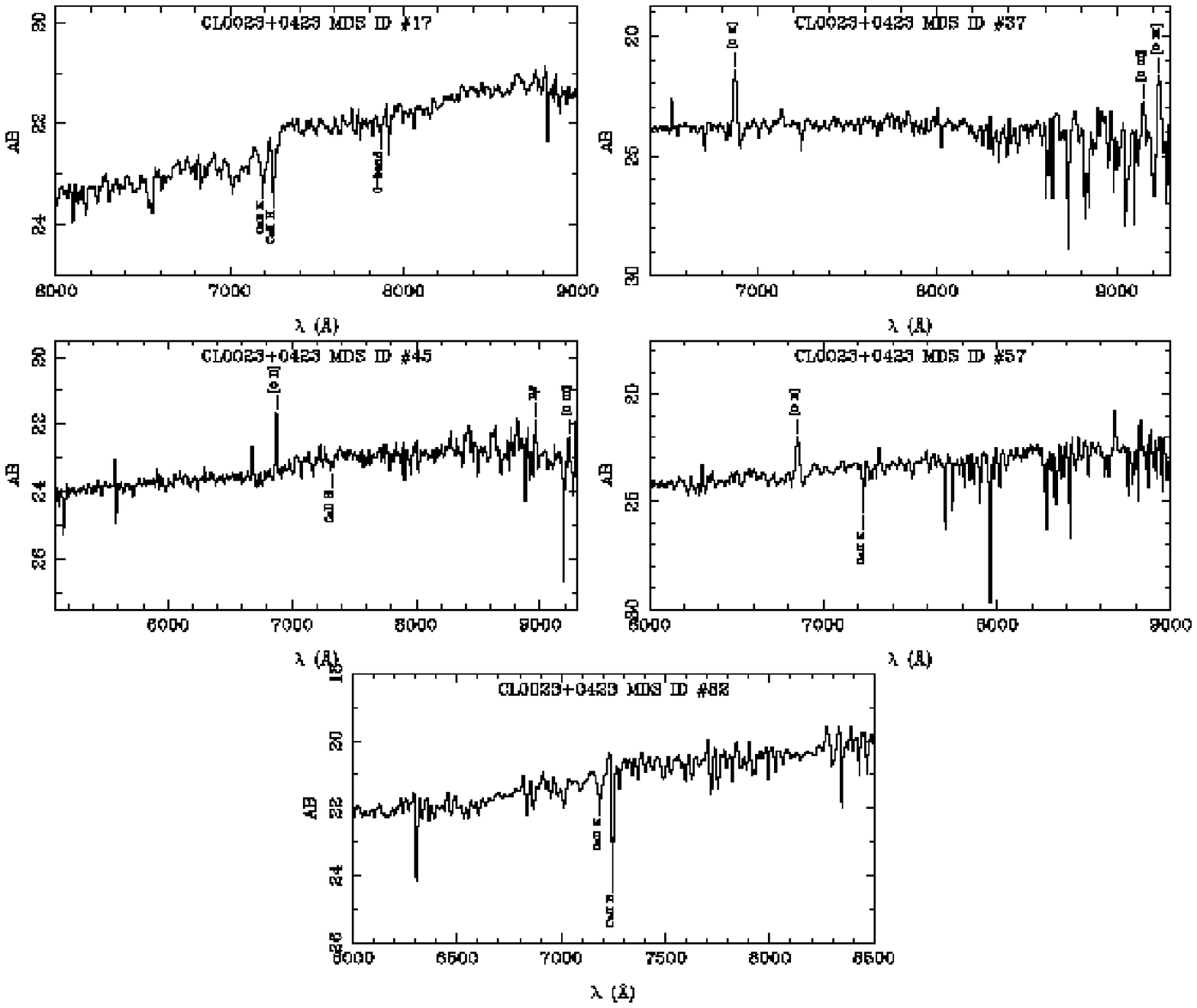}}
\caption{Spectra of five confirmed cluster members in
CL0023+0423 (see Sect.\ 5.3). These sample spectra include a classic
elliptical (MDS ID \#17), a disturbed spiral (MDS ID \# 37), and blue
compact galaxies (MDS ID \# 45 and 57). The spectra are untouched and,
therefore, reveal some of the difficulties associated with faint
object spectroscopy. For example, in some cases, poor sky subtraction
leaves obvious residual sky lines at 5577 \AA, 5891 \AA, and 6300,6363
\AA\ in the blue end of the spectrum and at $\simgreat 8000 \AA$ in
the red end of the spectrum (e.g\ MDS ID \# 45). In addition,
identified lines in the near-infrared which may seem unconvincing due
to the large number of residual sky lines in this region are actually
obvious in the two-dimensional spectrum (e.g\ MDS ID \# 37 and
45). The details of the line identification and redshift determination
are discussed in Sect.\ 4.2.1 of Paper I.}
\label{00spcluster}
\end{figure}

\begin{figure}
\centerline{\epsfbox{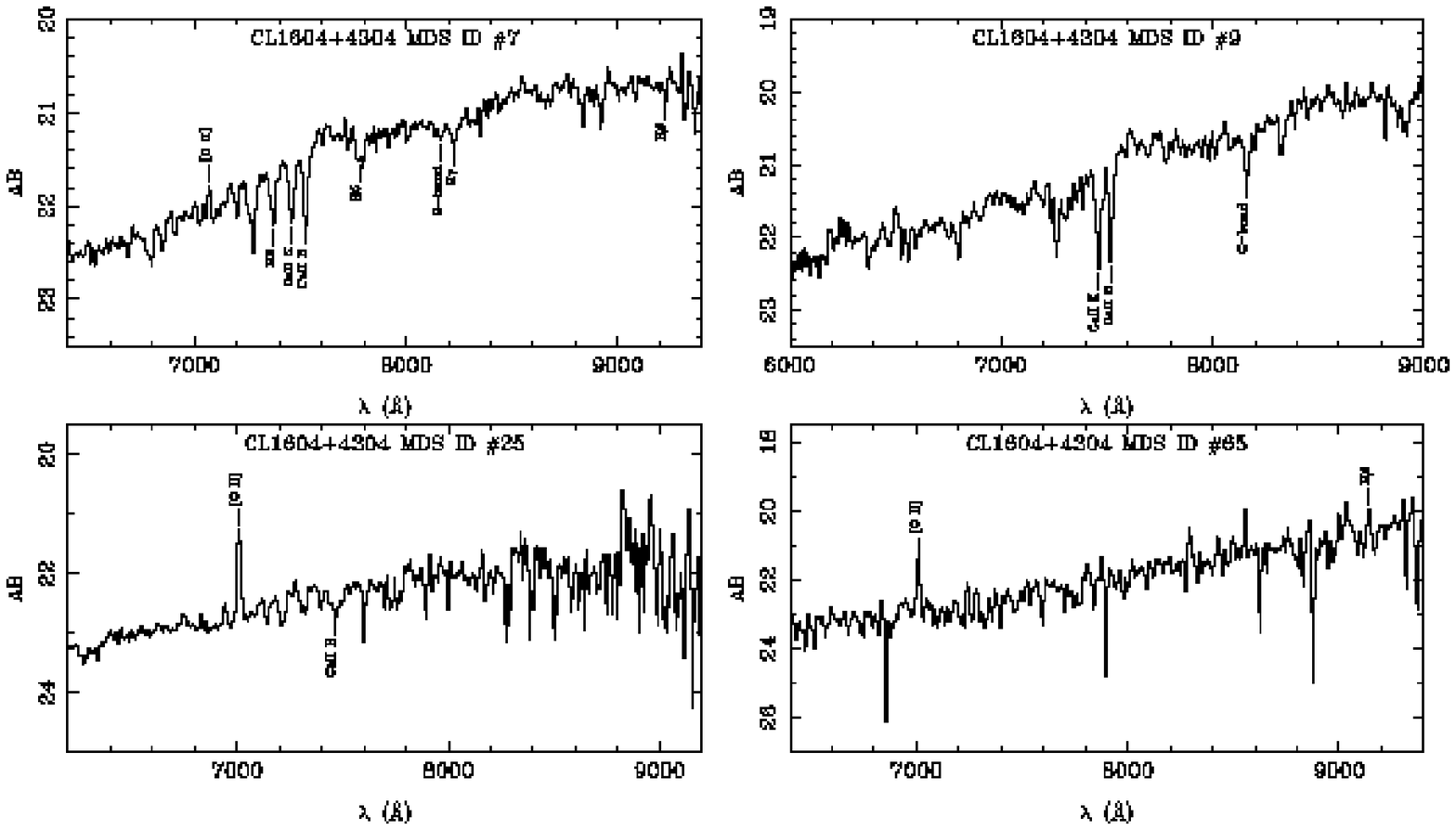}}
\caption{Spectra of four confirmed cluster members in CL1604+4304 (see
Sect.\ 5.3). These samples include spectra of an E+A galaxy (MDS ID
\# 7), a classic elliptical (MDS ID \# 9), a irregular/peculiarr
(MDS ID \#25), and an S0 (MDS ID \# 65). The spectra are untouched
and, therefore, reveal some of the difficulties associated with faint
object spectroscopy. For example, in some cases, poor sky subtraction
leaves obvious residual sky lines at 5577 \AA, 5891 \AA, and 6300,6363
\AA\ in the blue end of the spectrum and at $\simgreat 8000 \AA$ in
the red end of the spectrum (e.g\ MDS ID \# 25). In addition,
identified lines in the near-infrared which may seem un convincing due
to the large number of residual sky lines in this region are actually
obvious in the two-dimensional spectrum (e.g\ MDS ID \# 25 and
65). The details of the line identification and redshift determination
are discussed in Sect.\ 4.2.1 of Paper I.}
\label{16spcluster}
\end{figure}

\begin{figure}
\epsfysize=6.5in
\centerline{\epsfbox{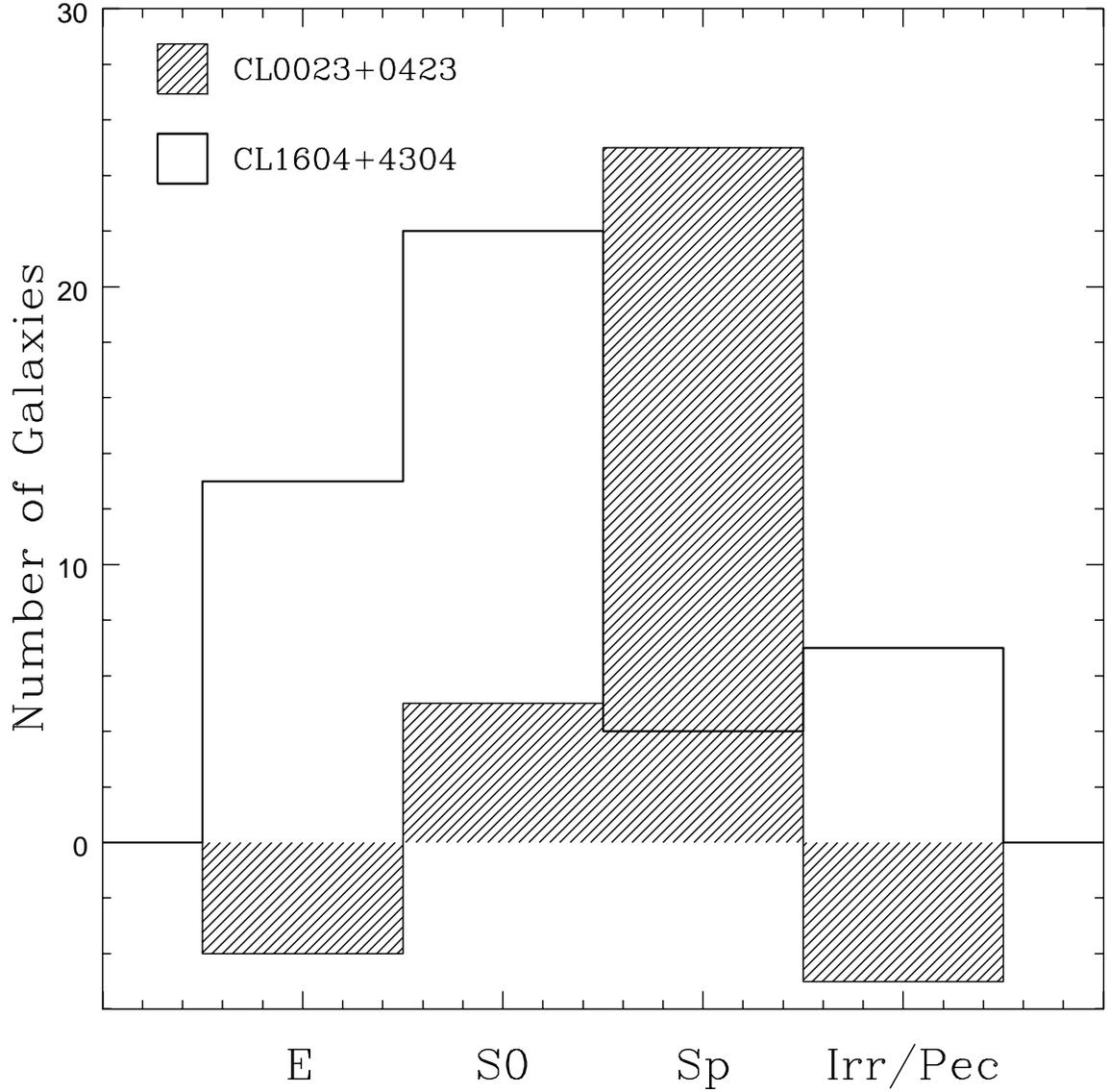}}
\caption{Distribution of galaxy morphology, brighter than $M_{V} = -19.0 +
5~{\rm log}~h$, in the two cluster fields. CL0023+0423 is indicated by
the shaded histogram, while CL1604+4304 is indicated by the solid-line
histogram. The distributions have been corrected for field
contamination using the morphologically classified counts from the MDS
and HDF (see Sect.\ 5.4).}
\label{morphfrac}
\end{figure}

%
% END 
% 

\end{document}